\documentclass[twocolumn]{aastex6}
\RequirePackage{lineno}
\usepackage{amsmath}
\usepackage{amssymb}
\usepackage{amsthm}
\usepackage{natbib,aasdefs,url,bm}
\usepackage{array}
\usepackage{float}
\usepackage{graphicx}
\usepackage{subfigure}
\usepackage{color}
\usepackage{mathtools,  epsfig, lipsum}

\newcommand\T{\rule{0pt}{2.6ex}}       % Top strut
\newcommand\B{\rule[-1.2ex]{0pt}{0pt}} % Bottom strut

\def\beq{\begin{equation}}
\def\eeq{\end{equation}}
\def\bey{\begin{eqnarray}}
\def\eey{\end{eqnarray}}

\begin{document}

\title{High-Energy Neutrinos from Millisecond Magnetars formed from \\ the Merger of Binary Neutron Stars}
 \author {Ke Fang \altaffilmark{1,2} \& Brian D.~Metzger\altaffilmark{3}}
 
\altaffiltext{1}{Department of Astronomy,  University of Maryland, College Park, MD, 20742, USA}
\altaffiltext{2}{Joint Space-Science Institute, University of Maryland, College Park, MD, 20742, USA}
\altaffiltext{3}{Department of Physics and Columbia Astrophysics Laboratory, Columbia University, Pupin Hall, New York, NY, 10027, USA}

\begin{abstract}
The merger of a neutron star (NS) binary may result in the formation of a long-lived, or indefinitely stable, millisecond magnetar remnant surrounded by a low-mass ejecta shell.  A portion of the magnetar's prodigious rotational energy is deposited behind the ejecta in a pulsar wind nebula, powering luminous optical/X-ray emission for hours to days following the merger.  Ions in the pulsar wind may also be accelerated to ultra-high energies, providing a coincident source of high energy cosmic rays and neutrinos.  At early times, the cosmic rays experience strong synchrotron losses; however, after a day or so, pion production through photomeson interaction with thermal photons in the nebula comes to dominate, leading to efficient production of high-energy neutrinos.  After roughly a week, the density of background photons decreases sufficiently for cosmic rays to escape the source without secondary production.  These competing effects result in a neutrino light curve that peaks on a few day timescale near an energy of  $\sim10^{18}$~eV.  This signal may be detectable for individual mergers out to $\sim$ 10 (100) Mpc by current (next-generation) neutrino telescopes, providing clear evidence for a long-lived NS remnant, the presence of which may otherwise be challenging to identify from the gravitational waves alone.  Under the optimistic assumption that a sizable fraction of NS mergers produce long-lived magnetars, the cumulative cosmological neutrino background is estimated to be $\sim 10^{-9}-10^{-8}\,\rm GeV\,cm^{-2}\,s^{-1}\,sr^{-1}$ for a NS merger rate of $10^{-7}\,\rm Mpc^{-3}\,yr^{-1}$, overlapping with IceCube's current sensitivity and within the reach of next-generation neutrino telescopes.
 \end{abstract}

\section{Introduction}

% NS merger remanent 

The merger of binary neutron stars (NS) are believed to be important sources of gravitational waves (GW) to be detected by the advanced LIGO \citep{1992Sci...256..325A, 0264-9381-27-8-084006}, advanced VIRGO \citep{1748-0221-7-03-P03012}, and future gravitational wave detectors.  NS mergers are also an important, if not dominant, contributor of rapid neutron capture ($r$-process) nucleosynthesis in the universe \citep{Lattimer&Schramm76,Symbalisty&Schramm82,Eichler+89,Freiburghaus+99}.  

The final outcome of a NS merger depends sensitively on the nuclear density equation of state (EoS) (e.g.~\citealt{Shibata&Taniguchi06}).  If the mass of the NS binary is high and/or if the EoS is relatively soft, then the merger results in a massive NS remnant which usually collapses into a black hole on a relatively short timescale of tens of milliseconds or less (e.g.~\citealt{Bauswein&Stergioulas17}).  If, on the other hand, the EoS allows a relatively large maximum non-rotating NS mass ($\gtrsim 2.3-2.4\,M_\odot$), then the merged core will create a long-lived supramassive NS (supported from immediate collapse even by its solid body rotation) or an indefinitely stable NS which never collapses (e.g. \citealt{2006Sci...311.1127D, Gao&Fan06,Metzger+08b,2010ApJ...724L.199O, 2012MNRAS.419.1537B, 2013ApJ...771L..26G, 2013ApJ...778...66K, MP14, 2015ApJ...812...24F, 2016ARNPS..66...23F, 2016arXiv161009381M,Ciolfi+17,Piro+17}).  

Due to the large angular momentum of the initial binary, any stable remnant NS will be rotating rapidly with an initial spin period close to the centrifugal break-up limit of $P_i \sim 1$~ms.  The remnant will also possess an ultra-strong internal magnetic field $\gtrsim 10^{15}-10^{16}$ G, due rapid amplification of the initially weak field by Kelvin-Helmholtz instabilities during the merger (e.g.~\citealt{2006Sci...312..719P, 2013ApJ...769L..29Z,Kiuchi+15,2015ApJ...809...39G}) and the magneto-rotational instability (e.g.~\citealt{Siegel+13,Mosta+15,Guilet+16,Radice17}).  The magnetar may also acquire a strong external dipole magnetic field $B \gtrsim 10^{13}-10^{15}$ G through magnetic buoyancy instabilities or via an efficient helical dynamo in the convective proto-magnetar \citep{1992ApJ...392L...9D}.  

A nearly maximally-spinning NS contains an enormous rotational energy reservoir of $\sim 10^{52}-10^{53}$ erg.  A strong dipole magnetic field provides a mechanism to extract this energy in the form of a powerful magnetized wind (e.g.~\citealt{Thompson+04}).  This wind emerges into the initially dense environment created by matter ejected promptly during the merger itself (e.g., \citealt{Hotokezaka+13,Sekiguchi+16}) and from the remnant accretion disk on a timescale of seconds (e.g.~\citealt{Fernandez&Metzger13,Metzger&Fernandez14,Just+15,Siegel&Metzger17}).  The total ejecta mass of both components is typically $M_{\rm ej}\sim 10^{-2}-10^{-1}\,M_\odot$ (e.g.~\citealt{Wu+16}).  At early times, this dense environment may collimate the magnetar wind into a bipolar jet  \citep{2012MNRAS.419.1537B}, providing one explanation for the long-lived X-ray emission observed after some short gamma-ray bursts (GRB; \citealt{Metzger+08b,Rowlinson+13,Gompertz+15}).  However, at later times, as the magnetar spins down, the weakening jet may become less stable, in which case a large fraction of its energy will instead be dissipated behind the ejecta by reconnection and shocks in the form of a pulsar wind nebula (PWN) \citep{MP14,Siegel&Ciolfi16a,Siegel&Ciolfi16b}.

A millisecond magnetar also provides a promising site for accelerating particles to ultra-high energies \citep{1969PhRvL..22..728G,2016ApJ...826...97P}.  A magnetospheric voltage drop of magnitude $\Phi_{\rm mag}\sim \mu\,\left(2\,\pi/P\right)^2/c^2=1.3\times10^{21}\,B_{14}\,P_{-3}^{-2}\,\rm V$ is produced across the open field lines that extend beyond the light cylinder located at $R_{\rm lc} = c/\Omega$, where $\mu \sim B\,R_*^3$ is the magnetic dipole moment, $B=10^{14}\,B_{14}\,\rm G$ is the strength of the surface magnetic field, $R_*\sim 10$~km is the NS radius, and $P=10^{-3}\,P_{-3}$~s is the  spin period of the magnetar.  An ion that taps a moderate fraction of this potential drop will reach ultrahigh energy (UHE; $E\geq10^{18}$~eV;  \citealt{2000ApJ...533L.123B, Arons03}). Specifically, particle acceleration can occur as charged particles surf-ride in the magnetar wind with a velocity along the radial component of the electric field \citep{Arons03}, or by magnetic reconnection of the opposite open magnetic fluxes in the equatorial current sheet   \citep{2014ApJ...785L..33P, 2014ApJ...795L..22C, 2015MNRAS.448..606C, 2016SSRv..tmp...84C}, or later by the wind termination shock \citep{2015JCAP...07..016L, 2015JCAP...08..026K}.

Cosmic rays accelerated in the nascent magnetar nebula interact with ambient photons and baryons.  For millisecond magnetars formed in normal core-collapse supernovae, the massive baryon envelope of the exploding progenitor star destroys particles accelerated at early times, significantly impacting the spectrum, flux, and chemical composition of the UHE cosmic rays (UHECR) that leak out of the nebula \citep{2012ApJ...750..118F, 2013JCAP...03..010F}. The interaction between cosmic rays and their surroundings produces charged pions that decay into high-energy neutrinos \citep{Murase09, Gao:2013ka, 2014PhRvD..90j3005F, 2015JCAP...06..004F} as well as neutral pions that decay into gamma rays \citep{2015ApJ...805...82M}.  Magnetars formed in binary NS mergers are surrounded by orders of magnitude less mass than normal core collapse supernovae.  However, a significant radiation field is still present as the result of non-thermal photons (X-ray) emitted by relativistic $e^\pm$ pairs in the nebula and thermal photons (optical/UV) emitted by the ionized ejecta \citep{MP14}.  This radiation may lead to the production of TeV-PeV neutrinos and GeV photons, which follow the arrival of the gravitational wave chirp.

% IceCube, UHECR, Fermi results 
The first detection of high-energy neutrinos was recently reported by  the IceCube Observatory (\citealt{1242856}; \citealt{Halzen:2016ee} for a review).  A   diffuse flux of TeV-PeV neutrinos with an astrophysical origin is measured at the level of $\sim 10^{-8}\,\rm GeV^{-1}\,cm^{-2}\,s^{-1}\,sr^{-1}$ per-flavor \citep{2015arXiv151005223T}. No neutrino above 10 PeV has been detected \citep{2016PhRvL.117x1101A}, and  no point source has been found in the 7-year data \citep{2017ApJ...835..151A}. The origin of these neutrinos remains a mystery \citep{2015AIPC.1666d0006M, Halzen:2016ee}. 

Multi-messenger searches by IceCube over past joint observational periods with LIGO and Virgo found no significant coincident events \citep{2014PhRvD..90j2002A}. The follow-up searches of gravitational wave events with ANTARES and IceCube are consistent with the expected background signal \citep{2016PhRvD..93l2010A, 2017arXiv170306298A}. General studies suggest that a joint search by advanced detectors is possible, although the chance of detection highly depends on specific source properties \citep{2011PhRvL.107y1101B, 2012PhRvD..85j3004B}. A template for the flux and light curve of neutrino emissions from merger products is timely and crucial for future searches. 

% In this work..
In this paper, we investigate  the evolution of high-energy neutrino emission from a long-lived millisecond magnetar following a NS merger, by studying particle interaction with the thermal and non-thermal radiation field emitted by the magnetar nebula. We adopt the photon field model of \citet{MP14} which accounts for the evolution of the thermal and non-thermal radiation, and their coupling through absorption by the ejecta. Different from \citet{Gao:2013ka} which considers the secondary emission by particles accelerated in the shocked ejecta, we focus on a general scenario that cosmic rays accelerated in the magnetosphere confront a spherical hot nebula. We account for additional interaction and cooling channels of cosmic ray particles that were not considered in \citet{2016ApJ...826...97P},  which can crucially impact the high-energy emission of the merger remnant, as demonstrated by \citet{Murase09} in the case of magnetars formed in supernovae.  

The paper is organized as follows. In Section~\ref{sec:photonField} we describe the time evolution of the radiation of the magnetar nebula. The photon field is then used to calculate cosmic ray interaction in Section~\ref{sec:interaction}.  We present the neutrino production from an individual merger event, as well as the integrated background signal from all-sky events, in Section~\ref{sec:results}.  We discuss our results and conclude in Section~\ref{sec:discussion}.

\section{Radiation Background}\label{sec:photonField}
Magnetic spin-down is assumed to dissipate a large fraction of the magnetar's rotational energy to form a hot nebula behind the ejecta, similar to well-studied PWN like the Crab Nebula \citep{1984ApJ...283..694K} but in several ways more extreme.  The nebula is composed of copious electron/positron pairs and non-thermal optical/UV/X-ray/$\gamma-$ray photons, due to a  cascade of high-energy photons resulted from inverse Compton scattering of  soft photons in the background ($e+\gamma\rightarrow e+\gamma$) and synchrotron emission in the nebula magnetic field, as well as electrons resulted from pair production of up-scattered photons ($\gamma+\gamma\rightarrow e^++e^-$).  When the optical depth due to pair production is high at these early times (``compactness" parameter $\gg 1$), most gamma-rays produce pairs before they can escape the ejecta.  Depending on the albedo of the ionized inner side of the ejecta, lower energy non-thermal UV/X-ray photons are either reflected back into the nebula, or absorbed and thermalized (see \citealt{Metzger+14} for a detailed discussion). 

As the ejecta expands with time, the optical depths of the nebula and ejecta decrease.  Photons diffuse out of this environment, powering luminous optical/UV/X-ray emission (\citealt{Kotera+13,2013ApJ...776L..40Y,MP14,Siegel&Ciolfi16a,Siegel&Ciolfi16b,  2017arXiv170406276H}).  At later times, as the ejecta becomes optically thin, the efficiency for thermalization decreases and the emission will become dominated by hard non-thermal X-ray/gamma-ray emission with a decreasing luminosity following the declining pulsar spin-down power.

% comment on Albeto factor, ionization effect, checks from X-ray observations
% Murase+2015 provides more detailed calculations on gamma-ray production. The high-energy end emission might not be relevant since cosmic rays mostly interact with the lower energy but more abundant thermal photons
% a 40 Gauss field exists in nebula. Check synchrotron emission 

Following \citet{MP14}, the evolution of non-thermal radiation $E_{\rm nth}$ and thermal radiation $E_{\rm th}$ are approximately described by
\beq \label{eqn:dE_nthdt}
\frac{dE_{\rm nth}}{dt} = L_{\rm sd} -\frac{E_{\rm nth}}{R}\frac{dR}{dt} - (1 - {\cal{A}})\,\frac{E_{\rm nth}}{t_d^n}, 
\eeq
\beq \label{eqn:dE_thdt}
\frac{dE_{\rm th}}{dt} = (1 - {\cal{A}})\,\frac{E_{\rm nth}}{t_d^n} -\frac{E_{\rm th}}{R}\frac{dR}{dt} -\frac{E_{\rm th}}{t_d^{\rm ej}}. 
\eeq

The first term on the right hand side of equation~\ref{eqn:dE_nthdt} is the magnetar's dipole spin-down power\footnote{Although we adopt the vacuum dipole expression for the spin-down rate, the expression for a force-free wind with an arbitrary inclination angle between the rotation and magnetic axes is identical to within a normalization factor of a few \citep{Spitkovsky06}.} \citep{1969ApJ...157.1395O}:
\bey
L_{\rm sd} & =&\frac{4}{9}\frac{\mu^2\Omega^4}{c^3} \\ \nonumber
 &=& 2.6\times10^{47}\,B_{14}^2\,P_{i,-3}^{-4}\,\left(1+\frac{t}{t_{\rm sd}}\right)^{-2}\,{\rm erg\,s^{-1} }\\ \nonumber
& \underset{t \gg t_{\rm sd}}=& 2.5\times10^{46}\, B_{14}^{-2}\,t_{5.5}^{-2}\,{\rm erg\,s^{-1}} 
\eey
where  $P_{i}=10^{-3}\,P_{i,-3}$~s is the initial spin period and
\beq
t_{\rm sd} \equiv \frac{E_{\rm rot}}{L_{\rm sd, 0}} = 1.4\times10^5 \,P_{i,-3}^2\,B_{14}^{-2}\,\rm s,
\eeq
is the initial spin-down time, where 
\beq
E_{\rm rot} = \frac{1}{2} I \Omega_i^2 = 3.6\times10^{52}\,P_{i,-3}^{-2}\,\rm erg
\eeq
is the total rotational energy  of a pulsar with assumed moment of inertia $I=2\,M_*\,R_*^2\,/5=1.8\times 10^{45}\,\rm g\,cm^2$, radius $R_{*} = 10$ km, and $M_{*} = 2.3M_{\odot}$.
At $t = t_{5.5}\,10^{5.5}\,{\rm s}\gg t_{\rm sd}$, the spin-down luminosity scales as $\propto t^{-2}$ for a pulsar braking index of $3$ \citep{1969ApJ...157.1395O}. 

The energy deposited by magnetic spin-down is shared between three sinks: i) kinetic energy of the ejecta; ii) thermal emission; and iii) non-thermal emission.  The second term in eqn.~\ref{eqn:dE_nthdt} and \ref{eqn:dE_thdt} describes the $PdV\sim (E/V)dV$ work done by the nebula on the ejecta, which causes the kinetic energy and mean ejecta velocity $v$ to increase according to 
\beq
M_{\rm ej}v\frac{dv}{dt} = \frac{E_{\rm nth}}{R}\frac{dR}{dt} + \frac{E_{\rm th}}{R}\frac{dR}{dt} 
\label{eq:dvdt}
\eeq
and thus the mean ejecta radius to increase from its small initial value $R_0 \approx 100$ km according to
\beq
R = \int^{t} v dt' + R_0
\label{eq:R}
\eeq
As long as the spin-down time of the magnetar is sufficiently short (compared to the time required for the ejecta to become transparent), then to good approximation most of the rotational energy is used to accelerate the ejecta, in which case one has
\beq  
\beta = \frac{v}{c} = \frac{1}{c}\left(\frac{2\,\int_0^t\,L_{\rm sd}dt'}{M_{\rm ej}} + v_{\rm 0}^2\right)^{1/2} \underset{t \gg t_{\rm sd}}\approx 1\,M_{\rm ej,-2}^{-1/2}\,P_{i,-3}^{-1}
\label{eq:betaest}
\eeq
where $v_{\rm 0}\sim 0.1\,c$ is the initial ejecta velocity and 
\beq
R \approx 9.5\times10^{15}\, \beta t_{5.5}\,\,\,\,{\rm cm}.
\label{eq:Rest}
\eeq 
We neglect special relativistic effects on the ejecta speed, which is not a terrible approximation as long as $M_{\rm ej} \gtrsim 10^{-2}M_{\odot}$, given many other (much larger) uncertainties in the formulation. Although we solve eqs.~(\ref{eq:dvdt},\ref{eq:R}) in our full numerical calculations, we employ eqs.~(\ref{eq:betaest},\ref{eq:Rest}) in our analytic estimates.

As non-thermal UV/X-ray photons produced by the nebula reach the ejecta walls, a fraction $(1- {\cal {A}})$ experiences bound-free absorption and is ``reprocessed" to thermal radiation \citep{Metzger+14}.  This process is described by the second loss term in eqn.~\ref{eqn:dE_nthdt} and the source term in  eqn.~\ref{eqn:dE_thdt}, where $\cal{A}$ is the frequency-averaged ``albedo" of the ejecta walls.  For simplicity we assume ${\cal{A}}=0$ (perfectly absorbing walls), which may be a reasonable approximation at UV/soft X-ray frequencies given the expected ionization parameter at times of interest \citep{MP14}.  However, our results do not depend qualitatively on this assumption as long as $\cal{A} \ne$ 1, and this approximation can be improved by a more detailed calculation of the ejecta ionization structure in future work.
 
The timescale for a photon to diffuse from the center of the nebula of size $\sim R$ to the inner edge of the ejecta (where it is absorbed or reflected) is given by
\beq
t_d^n \approx \frac{R}{c}\left(1+\tau_{\rm es}^n\right),
\label{eq:tdn}
\eeq
where $\tau_{\rm es}^n = n_{\pm}\sigma_T\,R $ is the Thomson optical depth across the nebula, $n_\pm$  is the pair density in the nebula, and $\sigma_T$ is  the Thomson cross section.   At early times, the high compactness parameter of the PWN, $\ell= E_{\rm nth}\,\sigma_T\,R/(V\,m_e\,c^2)=15.3\,B_{14}^{-2}\,t_{5.5}^{-3}\,\beta^{-2}$, results in copious pair production from $\gamma\gamma$ interactions.  The pair number density $n_\pm$ is estimated by assuming a balance between the pair creation rate, $\dot{N_\pm^+} = YL_{\rm sd} / (m_e\,c^2)$, and the pair annihilation rate, $\dot{N_\pm^-} = (3/16)\,\sigma_T\,c\,n_\pm\,N_\pm$, where $Y\approx 0.1$ is the pair multiplicity in a saturated state \citep{1987MNRAS.227..403S}.\footnote{The pair cascade resides in a saturated state ($\ell\gg1$) until a time $\sim 10^6$~s for  $B\sim 10^{14}$~G, thereby encompassing most epochs of relevance in this paper.} The optical depth is then $\tau_{\rm es}^n = \left(4\,Y\,L_{\rm sd}\,\sigma_T/\pi\,m_e\,c^3\,R\right)^{1/2}\approx  3.2\,B_{14}^{-1}\beta^{-1/2}t_{5.5}^{-3/2}$.  

Radiation travels freely through the nebula (without experiencing significant adiabatic losses) when the diffusion time is less than the ejecta expansion timescale, $t = R/\beta c$.  From eq.~\ref{eq:tdn}, this occurs after a time 
\beq
t_{d,0}^n = 7.5\times10^5\,B_{14}^{-2/3}\beta^{1/3}\,\rm s, 
\eeq
or equivalently when $\tau_{\rm es}^n \le \beta^{-1}$.  The optical depth of the nebula can therefore be convenientily re-expressed as $\tau_{\rm es}^n = \beta^{-1}\,\left(t/t_{d,0}^n\right)^{-3/2}$. 

The fraction $\left(1-\cal{A}\right)$ of the non-thermal radiation which is absorbed by the ejecta and reprocess is described by the last term in  eqn.~\ref{eqn:dE_thdt}.  The photon diffusion time through the ejecta is given by
\beq
t_d^{\rm ej} \approx \frac{R}{c}\left(1+\tau_{\rm es}^{\rm ej} \right),
\eeq
where  $\tau_{\rm es}^{\rm ej} = {3M_{\rm ej}\kappa}/{4\pi R^2}$ is the optical depth and $\kappa\sim 0.2-1\,\rm cm^2\,g^{-1}$ is the scattering/line opacity of the ejecta at optical/UV frequencies, which depends on the composition and ionization state of the ejecta (e.g.~\citealt{2000ApJ...530..757P,Kasen+15,Wollaeger+17}).\footnote{Given substantial neutrino irradiation of the disk wind ejecta in the case of a long-lived NS remnant, most of the ejecta is composed of Fe-group nuclei or light $r$-process nuclei \citep{Metzger&Fernandez14,Lippuner+17}. } As in the case of the nebula, photons can freely escape the ejecta once $t_d^{\rm ej} \le R/\beta c$, i.e.~after a timescale
\beq
t_{d,0}^{\rm ej} = 3.3\times10^4\, M_{-2}^{1/2}\left(\frac{\kappa}{0.2\,\rm cm^2\,g^{-1}}\right)^{1/2}\beta^{-1/2}\,\rm s. 
\eeq
The optical depth $\tau_{\rm es}^{\rm ej}$ can be expressed as $\tau_{\rm es}^{\rm ej} =\beta^{-1}\, \left(t/t_{d,0}^{\rm ej}\right)^{-2}$.   

The dominant loss terms in equation (\ref{eqn:dE_nthdt}) change with time. At early times  ($t\ll t_{d,0}^{n\,(\rm ej)}$), the kinetic term dominates the energy loss, while at late times  ($t\gg t_{d,0}^{n\,(\rm ej)}$), non-thermal (thermal) emission is more important.  Assuming instantaneous balance between the loss and source terms in eqn.~\ref{eqn:dE_nthdt}, one obtains the following approximate solution  (\citealt{MP14}) 
\bey\label{eqn:Enth}
E_{\rm nth} &=& \begin{cases} 
L_{\rm sd}\,t & t \ll t_{d,0}^{n} \\
L_{\rm sd}\,t_{d,0}^n\,\left( {t}/{t_{d,0}^n}\right)^{-1/2}   & t\gg t_{d,0}^n\end{cases}   \\ 
E_{\rm th} &=&\begin{cases} 
L_{\rm sd}\,t_{d,0}^n\,\left(t/t_{d,0}^n\right)^{5/2}, & t \ll t_{d,0}^{\rm ej} \\  \nonumber 
L_{\rm sd}\,t\,t_d^{\rm ej}/t_d^{n}, & t_{d,0}^{\rm ej} \ll t\ll t_{d,0}^{\rm n} \\ 
L_{\rm sd}\,t_{d,0}^{\rm ej}\,\left(t/t_{d,0}^{\rm ej}\right)^{-1}, & t\gg t_{d,0}^n, \end{cases}  
\eey
where we have assumed that $\chi \equiv t_{\rm d,0}^{\rm n} / t_{\rm d,0}^{\rm ej} = 22.7\,B_{14}^{-2/3}\,\beta^{5/6}\,M_{-2}^{-1/2} > 1$.  
As shown in Section~\ref{sec:interaction}, the time interval $t_{d,0}^{\rm ej} \ll t \ll t_{d,0}^n$ is most relevant epoch to particle interaction, during which time we have
\bey
E_{\rm nth}&=& 6.1\times10^{51}\,B_{14}^{-2}\,t_{5.5}^{-1}\,\rm erg,\\ 
E_{\rm th} &=& 5.1\times10^{50}\,B_{14}^{-1}\,\beta^{1/2}\,t_{5.5}^{1/2}\,\rm erg.
\eey
These expressions will prove useful in our analytic estimates below.

\begin{figure}
\includegraphics[width=\linewidth] {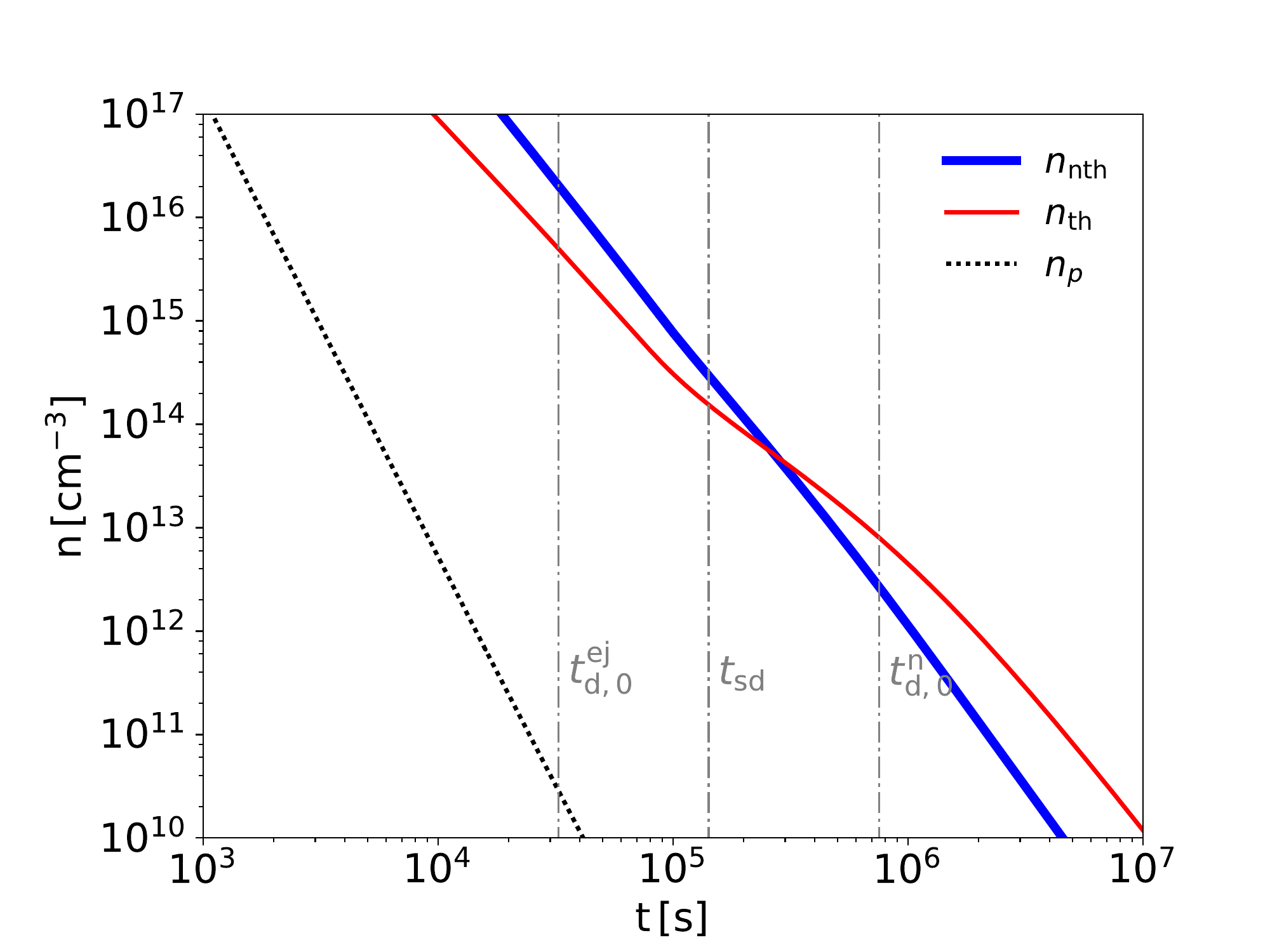}  
\caption{\label{fig:n_photon}
Number densities of thermal photons, non-thermal photons, and baryons of the ejecta, from a stable millisecond magnetar as a function of time since the merger.  Results are shown for a magnetar with initial rotation period $P=1\,\rm ms$ and surface magnetic field $B=10^{14}\,\rm G$. The photon number densities of non-thermal (thick blue line) and thermal (thin red line) radiation fields are computed by solving equations~\ref{eqn:dE_nthdt}, \ref{eqn:dE_thdt}, and \ref{eq:dvdt} numerically and  integrating photon energy distribution over all frequencies. The dotted black line shows the average number density of   baryons in the ejecta, computed from equation~\ref{eqn:np}, which is notably much lower than in the case of magnetars formed in supernovae.  Characteristic times are marked for reference with vertical lines, including the spin-down time $t_{\rm sd}$, and the time after which photons diffuse freely through the ejecta ($t_{d,0}^{\rm ej}$) and nebula ($t_{d,0}^n$) without substantial adiabatic losses.    }
\end{figure}

Thermal photons in the nebula have a temperature $T_{\rm th} =  \left({E_{\rm th}}/{V a}\right)^{1/4}$ and number density
\bey\label{eqn:nth}
n_{\rm th} &\sim & \frac{E_{\rm th}}{V\,k_B\,T_{\rm th}}  \approx  1.1\times10^{14}\,B_{14}^{-3/4}\,\beta^{-15/8}\,t_{5.5}^{-15/8}\,\rm cm^{-3} , \nonumber \\
\eey
where $a$ is the radiation constant.

We assume that non-thermal photons follow a spectrum $n(\varepsilon) \propto \varepsilon^{-2}$ from the thermal bath energy $\varepsilon_{\rm min}\sim 3\,k_B\,T_{\rm th}$ to the pair creation threshold $\varepsilon_{\rm max}\sim 2\,m_e c^2\sim 1\,\rm MeV$ \citep{1987MNRAS.227..403S}.  The number density can be estimated as
\bey\label{eqn:nnth}
n_{\rm nth} &\sim& \frac{E_{\rm nth}}{V\,\varepsilon_{\rm min}\,\ln\left(\varepsilon_{\rm max}/\varepsilon_{\rm min}\right)} \\ \nonumber
&=&3.3\times10^{13}\,B_{14}^{-7/4}\,t_{5.5}^{-27/8}\,\beta^{-19/8}\,\rm cm^{-3}.
\eey
The $\varepsilon^{-2}$ spectrum is motivated by observation of pulsar wind nebulae (e.g., \citealt{0004-637X-682-2-1166}).
In practice, the non-thermal spectrum will be more complicated than we have assumed. Relativistic leptons in the nebula can up-scatter both soft photons from the background (external inverse Compton) and the photons from their own synchrotron emission (synchrotron self-Compton). Depending on the acceleration mechanism, the intrinsic spectrum of leptons could  follow a broken power law.  More dedicated study taking into account these effects finds a $\gamma$-ray spectrum comparable or slightly softer than $\varepsilon^{-2}$ (e.g., \citealt{2015ApJ...805...82M}).  In our fiducial model, the density of thermal photons exceeds that of non-thermal photons at times most relevant to neutrino production, justifying moderate uncertainty in the non-thermal spectrum for our purposes.

The baryon density in the ejecta is given by
\beq\label{eqn:np}
n_p=\frac{M_{\rm ej}}{V\,m_p} = 4.2\times10^6\,M_{\rm ej,-2}\,t_{5.5}^{-3}\,\beta^{-3}\,\rm cm^{-3},
\eeq
which, due to the small ejecta mass and high velocity, is substantially lower than in the supernova case.

Figure~\ref{fig:n_photon} summarizes the number densities of the hadron and radiation backgrounds. Solid lines show the density of thermal (thin red) and non-thermal (thick blue) photons, as computed by  solving equations~\ref{eqn:dE_nthdt} and \ref{eqn:dE_thdt} and integrating the photon energy distribution over all frequencies.  Note that at times $\gtrsim 10^5$~s, the density of thermal photons exceeds that of non-thermal photons.  

Finally,  the nebula is strongly magnetized. The magnetic energy, $E_B = (B^2/8\,\pi)V$ evolves according to 
\bey
\frac{dE_B}{dt} = \epsilon_B\,L_{\rm sd} - \frac{E_B}{R}\,\frac{dR}{dt},
\eey
where the nebula magnetization $\epsilon_B\sim10^{-2}$ is motivated by observations of PWN such as the Crab Nebula (e.g, \citealt{1984ApJ...283..694K}) and the final term assumes the magnetic field is tangled and isotropic, such that it behaves effectively as a $\gamma = 4/3$ gas.  The magnetic field strength can be estimated by 
\beq\label{eqn:B_neb}
B_{\rm n} \approx \left(\frac{8\pi\epsilon_B L_{\rm sd}t}{V}\right)^{1/2}  \simeq 24.7\,\epsilon_{B,-2}^{1/2}\,B_{14}^{-1}\,\beta^{-3/2}\,t_{5.5}^{-2}\,\rm G. 
\eeq

\section{Particle Interaction}\label{sec:interaction}

\subsection{Particle Acceleration}
Ions extracted from the NS surface  gain energy by crossing open field lines  in the pulsar magnetosphere. A cosmic ray particle with charge Z can be accelerated to 
\bey\label{eqn:ECR}
E_{\rm CR} = \eta\,Z\,e\,\Phi_{\rm mag} = 4.1\times10^{19}\,Z\,\eta_{-1}\,t_{5.5}^{-1}\,B_{14}^{-1}\,\rm eV \nonumber \\ 
\eey
where $\eta=0.1\,\eta_{-1}$ is the acceleration efficiency, which can be interpreted as the fraction of the open-field voltage that  particles experience on average. 

The  charge density  demanded by the electromagnetic field, $\rho_{\rm GJ} =  -{\bf\Omega}\cdot{\bf B} / 2\pi c$ (the so-called Goldreich-Julian density; \citealt{JGR:JGR4198, Goldreich69}),  serves as a reasonable measure of  the ion  density \citep{Arons03}.  The cosmic ray production rate is  
\beq\label{eqn:dN_CRdt}
\dot{N} = \frac{ \rho_{\rm GJ}  } {Z e} \,2 A_{\rm pc} c = 8.5\times10^{37}\,B_{14}^{-1}\,Z^{-1}\,t_{5.5}^{-1}\,\rm s^{-1},
\eeq
where $A_{\rm pc} = \pi\,R_*^2\,\left(R_*/{R_{\rm lc}}\right)$ is the size of the polar cap. Combining eqns.~\ref{eqn:ECR} and \ref{eqn:dN_CRdt} we obtain the injection spectrum of cosmic rays:
\beq
\frac{dN}{dE} = \frac{9}{8}\frac{c^2\,I}{Z\,e\,\mu}\,\frac{1}{E} = 3.9\times10^{43}\,Z^{-1}\,E^{-1}\,B_{14}^{-1}.
\eeq
Notice that the pulsar spin-down results in a very hard $E^{-1}$ spectrum. The acceleration is expected to occur promptly as particles travel across the potential gap. 

\subsection{Interaction Rates of Cosmic Rays}

\begin{figure}[h]
\centering
\epsfig{file=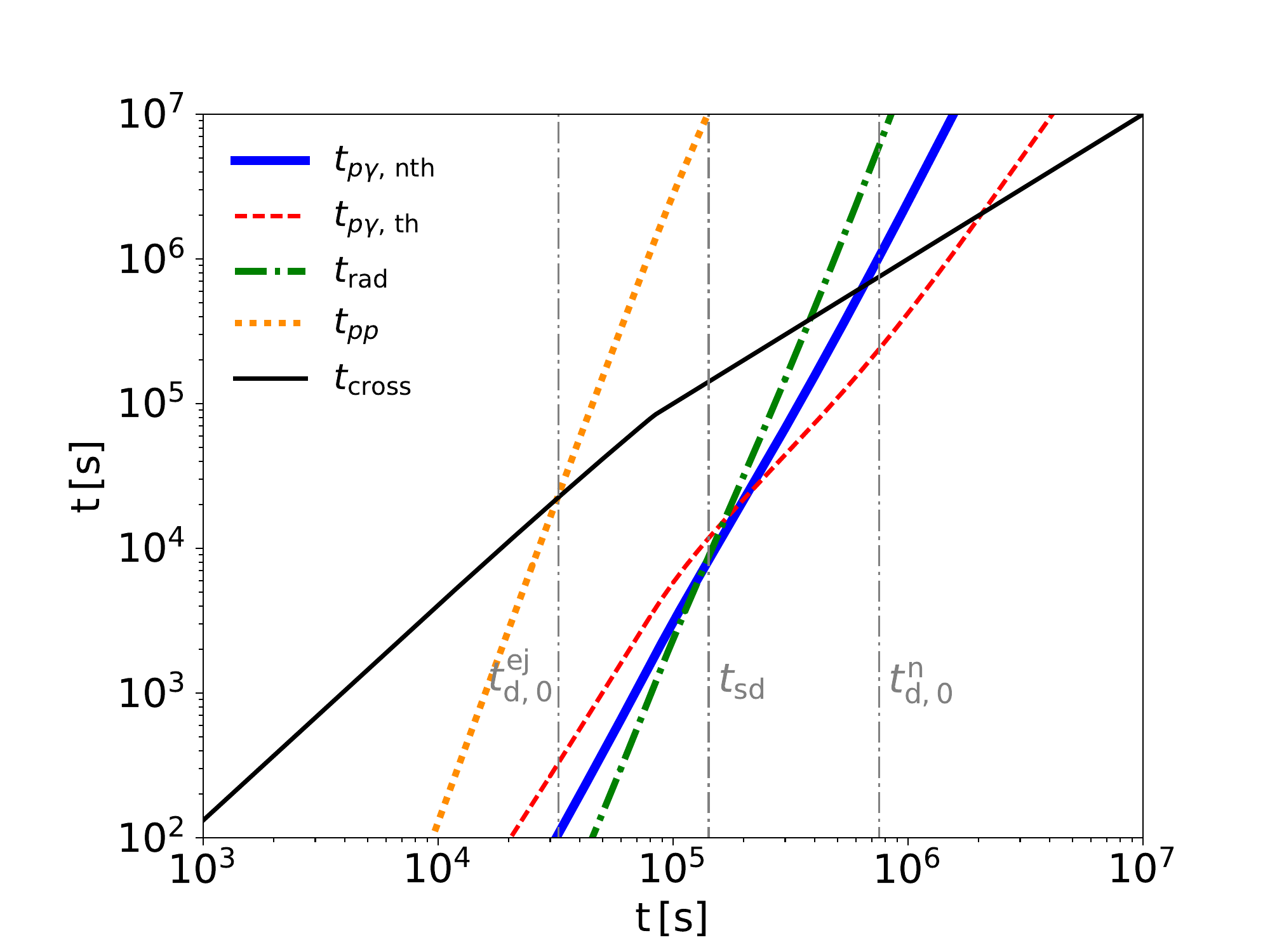,width=0.49\textwidth}  
\caption{\label{fig:t_proton}
Cooling timescales of cosmic ray protons as a function of time since the merger. Quantities shown include the time for photomeson interaction with  thermal (dashed red) and non-thermal (solid blue) radiation fields (equation~\ref{eqn:t_pgamma}), hadronuclear interaction with the ejecta baryons (dotted brown; equation~\ref{eqn:t_pp}), cooling due to the synchrotron emission (dash-dotted green; equation~\ref{eqn:t_p_IC}), and the light crossing time  $t_{\rm cross}$ (solid black).   }
\end{figure}

Accelerated particles possess Larmor radii $r_L =E/Z\,e\,B_n$ which are comparable to the total size of the nebula, $r_L/R = 0.6\,\eta_{-1}\,\beta^{1/2}\,\epsilon_{B,-2}^{-1/2}$. As the coherence length of a turbulent magnetic field is typically a fraction of the size of the magnetized region \citep{2009ApJ...705L..90C}, the particle Larmor radius is expected to be larger than the coherence length of the field.  Cosmic rays will therefore propagate in a semi-linear fashion through the nebula on the light crossing time $t_{\rm cross}\sim R/c = 10^{5.5}\,t_{5.5}\,\beta^{-1}\,\rm s$. 

During its propagation, a cosmic ray experiences three major cooling processes.  It  interacts with i) the nebular radiation via photonuclear  interaction,   and ii) ejecta baryons via hadronuclear interaction.  These two processes lead to the production of neutral and charged mesons, $p+\gamma (p)\rightarrow  p + \pi^{\pm,0}$ \citep{1990acr..book.....B}.  In addition, cosmic rays cool by radiative processes through iii) synchrotron radiation in the nebular magnetic field; this significantly suppresses the neutrino production at early times. Notice that  as $\sqrt{E_p\,\varepsilon_{\rm th}}\gg m_p\,c^2$, the inverse Compton process of an UHE proton is suppressed due to the Klein-Nishina effect and hence is negligible \citep{1979rpa..book.....R}. 

A proton with Lorentz factor $\gamma_p$ interacts with the photon field of spectrum $n(\varepsilon)=dn/d\varepsilon$ on a characteristic timescale given by
\bey\label{eqn:t_pgamma}
t_{p\gamma,\,\rm int}^{-1} = \frac{c}{2\gamma_p^2}\,\int_0^\infty\,d\varepsilon\,\frac{n(\varepsilon)}{\varepsilon^2}\,\int_0^{2\,\gamma_p\,\varepsilon}\,d\varepsilon'\,\varepsilon'\,\sigma_{p\gamma}(\varepsilon'),
\eey
where $\sigma_{p\gamma} $ is the cross section of  photopion production, which is $\sim 5\times10^{-28}\,\rm cm^2$ \citep{2004PhLB..592....1E}  at the $\Delta-$resonance, and $\sim1.6\times10^{-28}\,\rm cm^{-2}$ above the resonance. 
The cooling time   is  $t_{\,p\gamma} = t_{p\gamma,\,\rm int}/ \kappa_{p\gamma} $, where $\kappa_{p\gamma}\sim 0.15$  is the average fraction of energy lost from a proton per collision  (the  ``elasticity").   The  cooling time due to photomeson interaction with  non-thermal and thermal fields can be roughly estimated as $t_{p\gamma}\sim \left(n_\gamma\,\sigma_{p\gamma}\,\kappa_{p\gamma}\,c\right)^{-1}$, which for the thermal and non-thermal photon densities (eqs.~\ref{eqn:nth},\ref{eqn:nnth}) is given by
\bey\label{eqn:t_pg_nth}
t_{p\gamma,\,\rm th}  &=&   1.3\times10^{4}\,t_{5.5}^{15/8}\,B_{14}^{3/4}\,\beta^{15/8}\,\rm s \\ 
t_{p\gamma,\,\rm nth}  &=& 4.2\times10^{4}\,t_{5.5}^{27/8}\,B_{14}^{7/4}\,\beta^{19/8}\,\rm s
\eey

Due to the low ejecta density (eq.~\ref{eqn:np}), the timescale for hadronuclear interaction is by comparison  longer,
\beq \label{eqn:t_pp}
t_{pp} = \left(n_p\,\sigma_{pp}\,\kappa_{pp}\,c\right)^{-1} = 1.6\times10^8\,M_{\rm ej,-2}^{-1}\,t_{5.5}^3\,\beta^3\,\rm s,
\eeq
where $\sigma_{\rm pp}\sim 10^{-25}\,\rm cm^2$ (at around $10^{18}$~eV) and $\kappa_{\rm pp}\sim 0.5$ \citep{2004PhLB..592....1E}. 

Photopion and hadronuclear interactions lead to the creation of charged pions, which decay into neutrinos \citep{1990acr..book.....B}. The total pion creation rate is given by
\beq
t^{-1}_{\pi,\rm cre} = t_{p\gamma,\,{\rm th}}^{-1} +  t_{p\gamma,\,{\rm nth}}^{-1} + t_{pp}^{-1}
\eeq

Pion production must compete with synchrotron cooling of the proton, which occurs on a timescale
\bey\label{eqn:t_p_IC}
t_{p,\,\rm rad} &=&  \frac{3\,m_p^3\,c}{4\,\sigma_T\,m_e^2\,\gamma_p\,u_B} \\ \nonumber
&=&1.8\times10^5\,\eta_{-1}^{-1}\,t_{5.5}^5\,B_{14}^3\,\beta^3\,\epsilon_{B,-2}^{-1}\,\rm s,
\eey
where we have used equation (\ref{eqn:B_neb}) to estimate the nebular magnetic field.  Pion creation also effectively ceases once the formation timescale exceeds the age of the source, $t_{\pi,\,\rm cre} \ge t_{\rm cross}$.  

Combining the effects described above, pion creation is effectively suppressed by a factor 
\beq
f_{\rm sup}^p =  \max(1,  \frac{t_{\rm cross}}{t_{\pi,\rm cre}},  \frac{t_{\rm p,rad}}{t_{\pi,\rm cre}}). 
\eeq

Figure~\ref{fig:t_proton} compares the proton cooling time of different processes as a function of time since the merger.  At the earliest times, all cooling processes are much shorter than the nebula crossing time.  Radiative cooling (green line) dominates at early times, suppressing pion production.  However, this gives way to photopion interaction with thermal photons at $t \gtrsim 10^{5.5}$ s.  Then, at late times $t \gtrsim 2\times10^6\,\rm s$, the radiation field becomes too dilute to interact with protons accelerated by the pulsar, and the window of pion (and thus neutrino production) closes.

\subsection{Interaction Rates of Pions and Muons}
\begin{figure}[h]
\centering
\epsfig{file=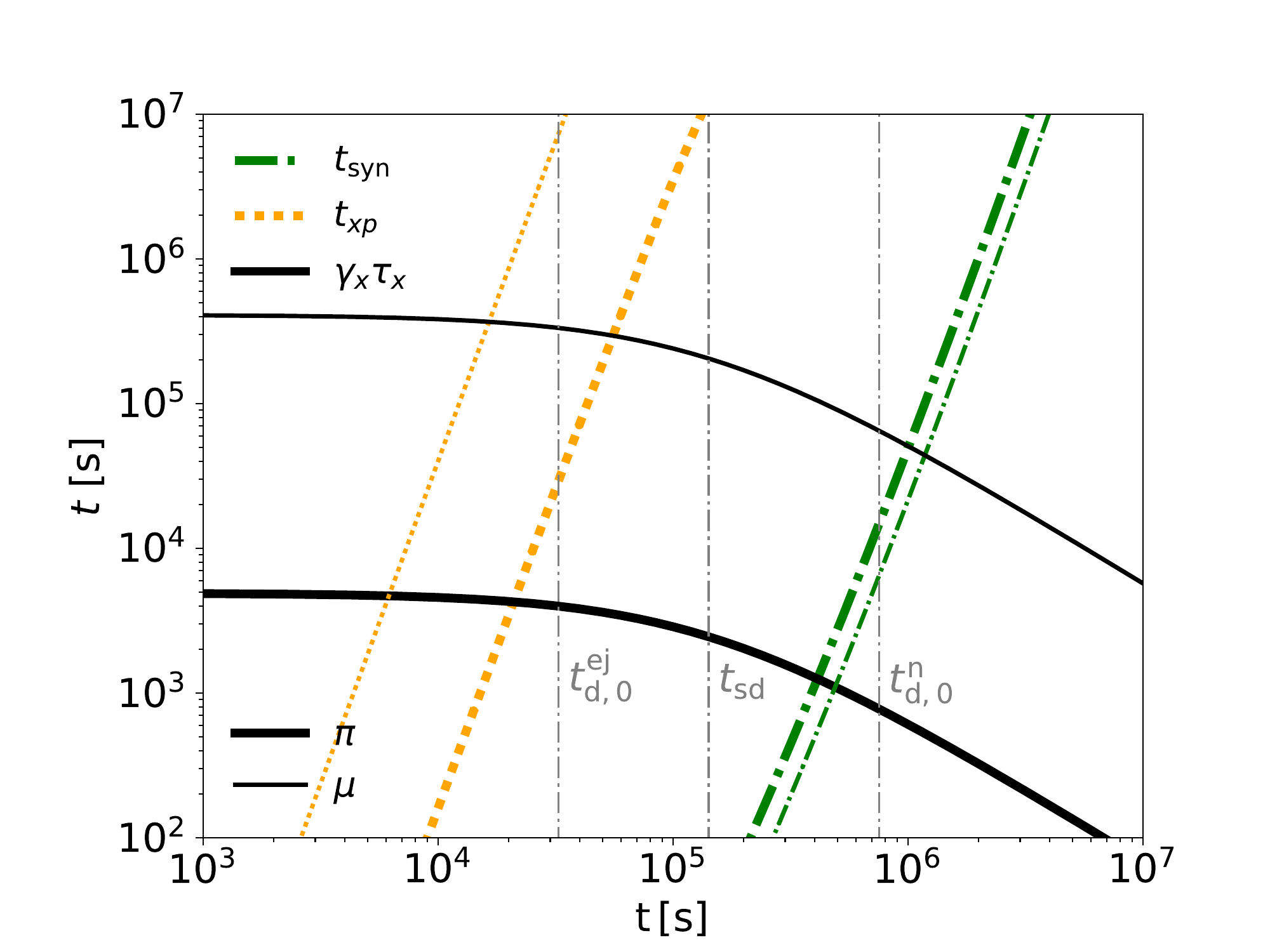,width=0.49\textwidth}  
\caption{\label{fig:t_pi}
Lifetime of pions (thick lines) and muons (thin lines) in the lab frame (solid black), compared to their characteristic cooling time due to hadronuclear interaction with the ejecta baryons (dotted brown; equation~\ref{eqn:t_xp}) and synchrotron radiation in the nebula (dash-dotted green; equation~\ref{eqn:t_x_rad}). }
\end{figure}

Charged mesons created by photopion and hadronuclear interactions decay into neutrinos via $\pi^\pm \rightarrow \mu^\pm + \nu_\mu(\bar{\nu}_\mu)\rightarrow e^\pm + \nu_e (\bar{\nu}_e)+\nu_\mu+\bar{\nu}_\mu$.  The neutrino production  competes with  the radiative and hadronic cooling of the mesons and muons. The latter occur at a rate
\beq
t_{x,\,\rm c}^{-1} = t_{x p}^{-1} + t_{x,\rm rad}^{-1}, 
\eeq
where $x$ denotes either $\pi$ or $\mu$,
\beq\label{eqn:t_xp}
t_{x p} = \left(n_p\,\sigma_{x p}\,\kappa_{x p}\,c\right)^{-1}
\eeq
is the hadronic cooling rate due to interaction with the ejecta baryons, and 
\bey \label{eqn:t_x_rad}
t_{x,\,\rm rad} = \frac{3\,m_x^4\,c^3}{4\,\sigma_T\,m_e^2\,E_x\,u_B}  
\eey
is the energy loss time due to synchrotron radiation. 
The relevant time scales for pions and muons are shown in Fig.~\ref{fig:t_pi}. Synchrotron emission dominates the energy loss until $\sim 10^{5.5}$~s for pions and $\sim10^6$~s for muons.

These cooling processes can be accounted for by introducing a second suppression factor on the neutrino production rate of the form,
\beq
f_{\rm sup}^x = \min \left(1, \frac{t_{x,\rm c}} {\gamma_x\,\tau_x} \right)
\eeq
This quantifies the fact that neutrinos are efficiently produced only if the decay time of a pion or muon is shorter than its cooling time. 

The suppression factor can be estimated analytically as
\bey\label{eqn:t_pi}
f_{\rm sup}^\pi &=& 0.3 \, \eta_{-1}^{-2}\,B_{14}^4\,\beta^3\,\epsilon_{B,-2}^{-1}\,t_{5.5}^6\\
f_{\rm sup}^\mu &=& 1.5\times10^{-3}\, \eta_{-1}^{-2}\,B_{14}^4\,\beta^3\,\epsilon_{B,-2}^{-1}\,t_{5.5}^6
\eey
where $\sigma_{\pi p} = 5\times10^{-26}\,\rm cm^2$, $\kappa_{\pi p}\sim0.8$, $\tau_\pi =2.6\times10^{-8}\,\rm s$, $\sigma_{\mu p} = 2\times10^{-28}\,\rm cm^2$, $\tau_\mu =2.2\times10^{-6}\,\rm s$ \citep{2004PhLB..592....1E}, and taking $E_\pi\sim 0.2\,E_p$ as the average ratio of pion energy to its parent proton energy in photopion production.  Because the mean lifetime of a muon exceeds that of a pion by a factor of $\sim$100, muons almost immediately experience radiative cooling before decaying into  secondary neutrinos.

\section{Neutrino Production}\label{sec:results}
\subsection{Individual sources}

\begin{figure}[h!]
\centering
\epsfig{file=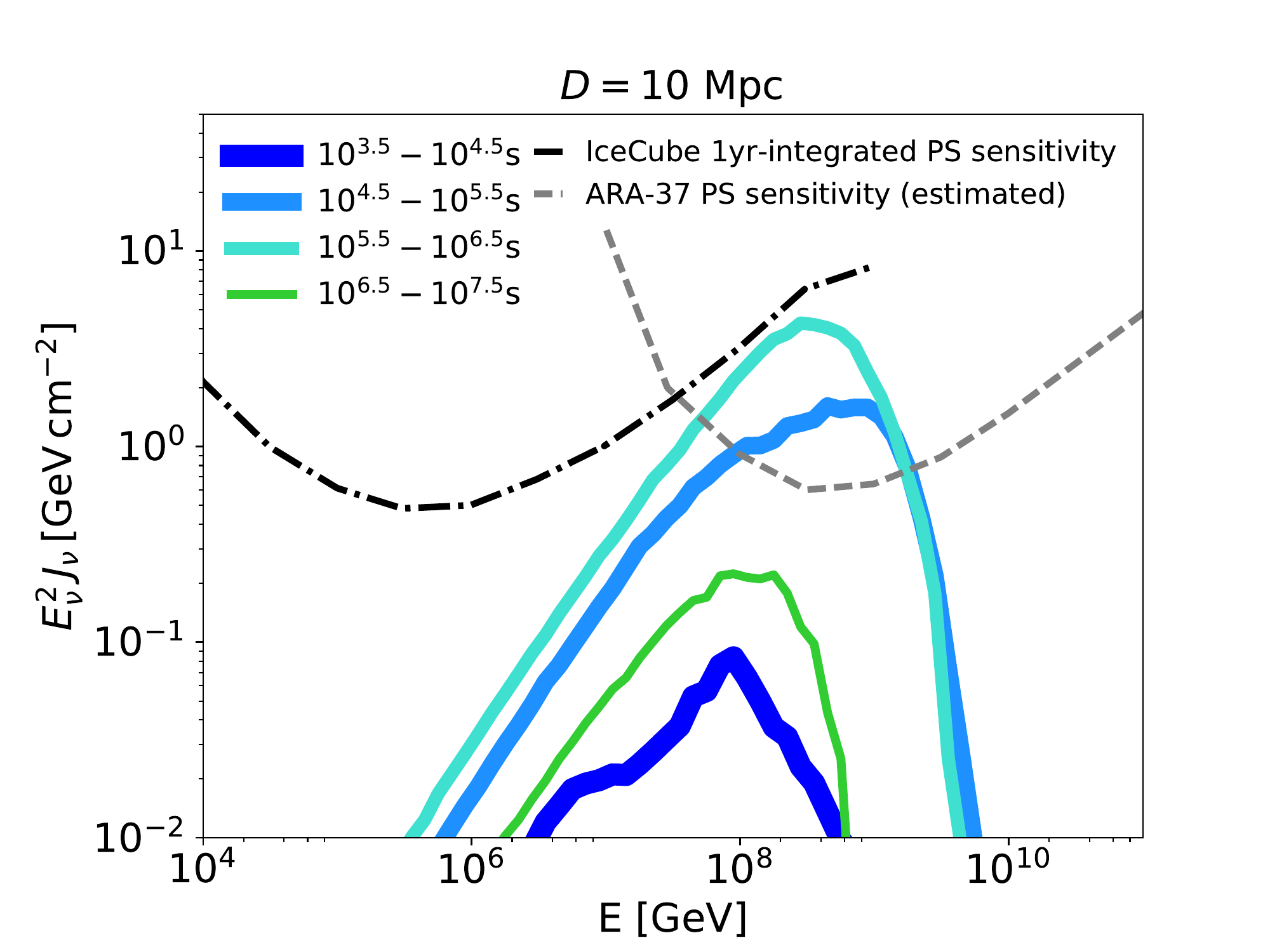,width=0.49\textwidth}  
\caption{\label{fig:fluence}
All-flavor fluence of high-energy neutrinos from a stable millisecond magnetar on timescales from an hour to a year (solid lines) after the merger.   The fiducial magnetar model assumes an initial spin period $P_{i} = 1$ ms, surface dipole magnetic field $B=10^{14}$ G, ejecta mass $M_{\rm ej}=0.01\,M_\odot$, and source distance $D = 10$ Mpc. The black dash-dotted line indicates the  90\% sensitivity of IceCube for a time-integrated search of point-like sources with one year of operation \citep{2017ApJ...835..151A} (which is comparable to its time-dependent sensitivity for a transient source with week-long duration; \citealt{Aartsen:2015jx}).  The grey dashed line shows the estimated point-source sensitivity of ARA \citep{2012APh....35..457A} (or ARIANNA; \citealt{2015APh....70...12B}) from an one-year time-integrated search.   }
\end{figure}

\begin{table*}[t]
\caption{Summary of characteristic timescales} \label{table:timescales}
\centering
\begin{tabular}{ccc}
\hline\hline 
time symbol \T & value [s] at $t_{\rm sd} \ll t \ll t_{d,0}^n$ & description \B  \\
\hline
$t_{d,0}^{\rm ej} \T $ & $3.3\times10^4\, M_{-2}^{1/2}\,\beta^{-1/2}$  & time after which photons escape the ejecta freely \\
$t_{\rm sd}$ & $1.4\times10^5 \,P_{i,-3}^2\,B_{14}^{-2}$ & spin-down time \\
$t_{\rm sup,\,0}^p$  & $1.4\times10^5\,\eta_{-1}^{8/25}\,B_{14}^{-18/25}\,\beta^{-9/25}\,\epsilon_{B,-2}^{8/25}$ & time when an efficient photopion production starts   \\
$t_{\rm sup,\,0}^\pi$  & $3.9\times10^5\,\eta_{-1}^{1/3}\,B_{14}^{-2/3}\,\beta^{-1/2}\,\epsilon_{B,-2}^{1/6}$ & time when $\pi$s decay  \\
$t_{d,0}^{n} \T $ & $7.5\times10^5\,B_{14}^{-2/3}\beta^{1/3}$ & time after which photons escape the nebula freely \\
$t_{\rm sup,\,0}^\mu$  & $9.3\times10^5\,\eta_{-1}^{1/3}\,B_{14}^{-2/3}\,\beta^{-1/2}\,\epsilon_{B,-2}^{1/6}$ & time when $\mu$s decay  \\
$t_{\pi, 0} $ \T& $1.2\times10^7\,\beta^{-23/7}\,B_{14}^{-6/7}$ & time when neutrino production stops \B \\
\hline
\end{tabular}
\end{table*}

Neutrino production is delayed until charged pions are both produced efficiently and avoid being cooled radiatively before decaying.  The former occurs first, after the pion production rate exceeds the proton cooling timescale once $t_{\rm sup,\,0}^p \equiv t\left(t_{p,\,\rm rad} = t_{\pi,\,\rm cre}\right)$.  However, radiative cooling of the pions prevents neutrino production until somewhat later, once $t_{\rm sup,\,0}^\pi \equiv t\left(t_{\pi,\,\rm rad} = \gamma_\pi\,\tau_\pi\right)$.  At yet later times, muons obey the same decay timescale condition and thus also contribute to neutrino production; this happens after $t_{\rm sup,\,0}^\mu \equiv  t\left(t_{\mu,\,\rm rad} = \gamma_\mu\,\tau_\mu\right)$.  Neutrino production effectively ceases once the creation process freezes out, as occurs after the time $t_{\pi,0}\equiv t\left (t_{\pi,\,\rm cre} = t_{\rm cross}\right)$.  Substantial high energy neutrino production is therefore typically limited to a window of hours to weeks following the merger.  These characteristic timescales, as well as other important timescales in the problem, are summarized in Table~\ref{table:timescales}. 

The neutrino flux can be estimated by
\bey
E_\nu^2\,\frac{dN_\nu}{dE_\nu} = E_{\rm CR}^2\,\frac{dN_{\rm CR}}{dE_{\rm CR}} \,\frac{f_{\rm sup}^p}{2}\,f_{\rm sup}^\pi\left(\frac{1}{4} + \frac{1}{2}\,f_{\rm sup}^\mu\right).\nonumber \\
\eey
The first $1/2$ factor results because charged pions are produced in a $p\gamma$ interaction only half the time on average.  The final factor in parentheses results from the fact that, when a pion decays, $\sim1/4$ of its energy goes to a muon neutrino; by contrast, when a muon decays, $\sim 2/3$ of its products are neutrinos. 

The neutrino flux peaks near the time $t_{\rm sup,0}^\mu$.  For a fiducial parameters, a source at distance $D = 10\,\rm Mpc$ produces a peak fluence given by
\bey\label{eqn:fluence}
\left(E_\nu^2\,J_\nu\right)_{\rm peak} &\approx& \left.\frac{E_\nu^2}{4\,\pi\,D^2}\frac{dN_\nu}{dE_\nu}\right|_{t= t_{\rm sup,\,0}^\mu}\\ \nonumber
&=& 9.0\,\eta_{-1}^{2/3}\,B_{14}^{-4/3}\,\beta^{1/2}\,\epsilon_{B,-2}^{-1/6}\, D_{10\,\rm Mpc}^{-2}\\ \nonumber
&& f_{\rm sup}^p \,f_{\rm sup}^\pi\,  f_{\rm sup}^\mu\,\rm GeV\,cm^{-2}. 
\eey

To account in greater detail for the energy distribution of pions from a photopion production, we calculate the neutrino flux semi-analytically using the numerical package SOPHIA \citep{2000CoPhC.124..290M}. 
At each time step, we calculate the energy and flux of cosmic ray protons injected from the pulsar magnetosphere according to eqs.~\ref{eqn:ECR} and \ref{eqn:dN_CRdt}. These cosmic rays meet non-thermal and thermal photons in the nebula with number densities and spectra as determined by solving equations~\ref{eqn:dE_nthdt} and \ref{eqn:dE_thdt}.  Meanwhile, protons are cooled by synchrotron radiation in the nebula magnetic field (equation~\ref{eqn:B_neb}). A fraction $f_{\rm sup}^p$ of the injected protons undergo photopion production. We use SOPHIA to compute the pion produced by UHECR interaction in the thermal background. Each $\pi^\pm$ product, depending on its energy $E_\pi$ and the system time, contributes a number of $f_{\rm sup}^\pi$ neutrino with energy $E_\pi/4$, and the same number of muon with energy $3\,E_\pi/4$. Each muon, again depending on its energy $E_\mu$ and the system time, contributes $2\times f_{\rm sup}^\mu$ neutrinos each with energy $E_\mu/3$.  

The neutrino flux from a stable fast-spinning magnetar formed from the merger is shown in Fig.~\ref{fig:fluence}. For initial spin period $P_i = 1$ ms and surface magnetic field $B=10^{14}$ G, the neutrino emission starts $\sim 1$~h after the merger, reaches the peak after $\sim 4$~days, and lasts for about a year before decreasing to $<5\%$ of the maximum flux. The neutrino spectrum peaks near an energy of $10^{17.5}$ eV. 

For comparison, we show the time-integrated sensitivity of IceCube  with full configuration and one year operation for a source in the declination band $0<\delta<30^\circ$ \citep{2017ApJ...835..151A}. Note that this sensitivity is comparable to the IceCube sensitivity from a time-dependent search for a flare with week-long duration \citep{Aartsen:2015jx}. With millisecond initial spin period and $10^{14}$ G surface magnetic field, a post-merger magnetar is detectable by IceCube only when it is within $\sim10\,\rm Mpc$.
With sensitivity windows focusing on EeV energies, the projected Askaryan Radio Array (ARA, \citealp{2012APh....35..457A}) and the Antarctic Ross Ice Shelf Antenna Neutrino Array (ARIANNA \citealp{2015APh....70...12B}) will be promising detectors to observe these neutrinos. In Fig.~\ref{fig:fluence}, we estimate the point-source sensitivity of ARA/ARIANNA by scaling the IceCube point-source sensitivity by the ratio of the differential sensitivity of ARA/ARIANNA to that of IceCube at 1~EeV. Such a crude estimation assumes that ARA/ARIANNA has an angular resolution that is comparable to that of IceCube at EeV, and that its effective area is independent of energy.\footnote{ARA/ARIANNA would improve the angular resolution of IceCube by a factor of $\sim 2$ and the effective area by a factor of 5 over two decades in energy \citep{2012APh....35..457A}.  The shape and flux level of an ARA/ARIANNA sensitivity curve from the collaboration(s) in the future could therefore be different from our estimates in Figure~\ref{fig:fluence}.}
Next generation telescopes, such as IceCube-Gen2 \citep{2014arXiv1412.5106I}, giant radio array for neutrino detection (GRAND, \citealp{2017EPJWC.13502001M}), and Cherenkov from astrophysical neutrinos telescope (CHANT, \citealp{2017PhRvD..95b3004N}) are expected to offer even better sensitivities that can probe high-energy neutrinos emitted by more distant mergers.    

Secondary protons and neutrons from the photomeson production may continue to participate in higher-order interactions, resulting in additional neutrinos at low energies \citep{2015JCAP...06..004F}. Moreover, UHE photons from the decay of neutral pions may cascade in or escape from the nebula, leading to potential observational signatures for nearby sources. The higher-order effect and UHE photon production will be explored in future work. 

\subsection{Diffuse flux}

\begin{figure}
\centering
\epsfig{file=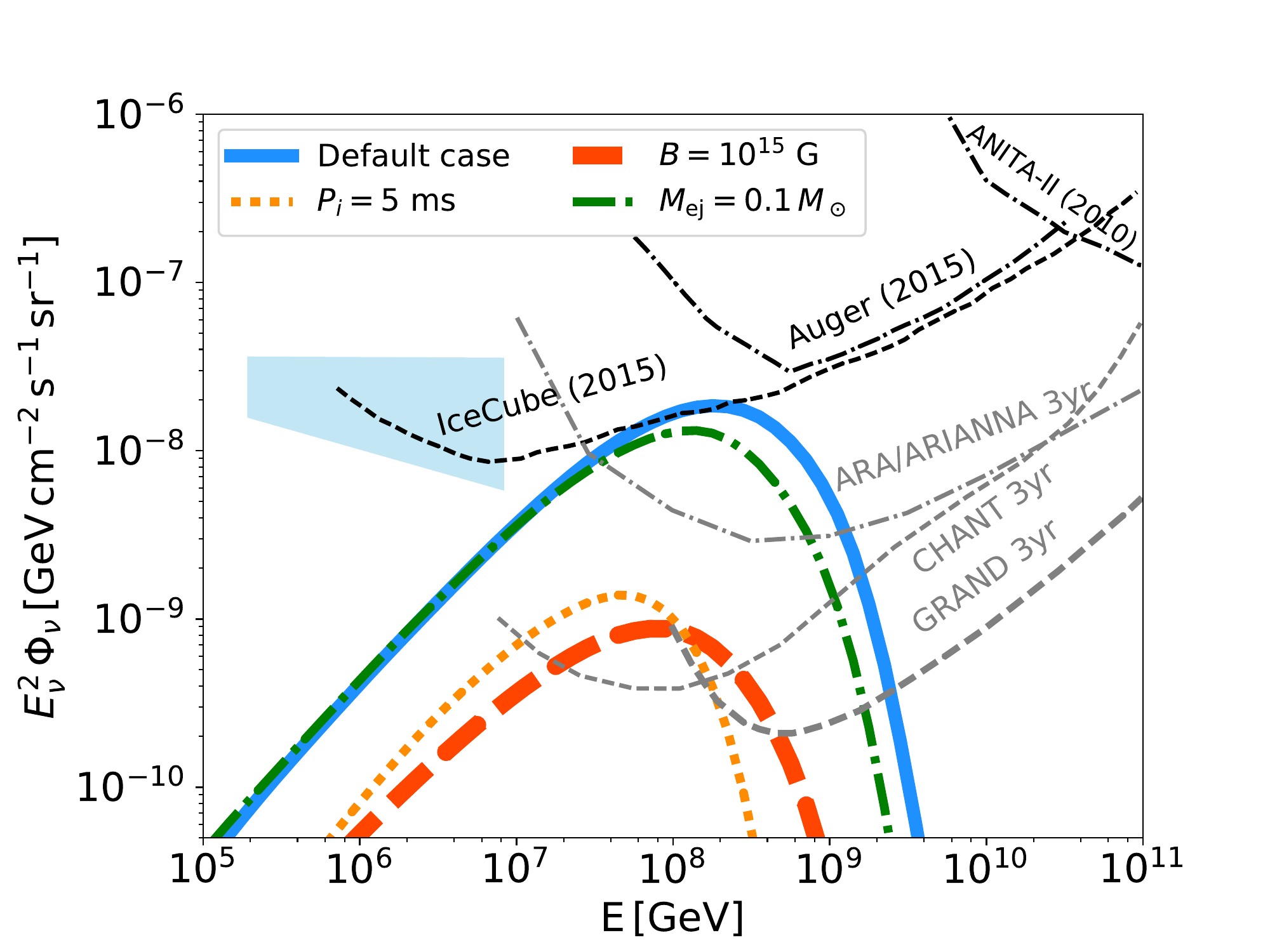,width=0.49\textwidth}  
\caption{\label{fig:fluence_multi}
All-flavor diffuse neutrino spectra for magnetar populations with different $B$, $P_i$, or $M_{\rm ej}$ values as labelled with different colors.  All models assume that a large fraction of mergers result in the formation of a long-lived or stable magnetar ($f_{\rm mag} = 1$) and adopt a local NS merger rate of ${\cal R}(0)\sim 10^{-7}\,\rm Mpc^{-3}\,yr^{-1}$.  In each case, other than the indicated parameter, the other parameters are set to be their default values as in Figure~\ref{fig:fluence}.  More magnetized or slower spinning NSs are more challenging to detect. Heavier ejecta mass expands less rapidly and produces less neutrinos at early times due to more severe radiation cooling of primary and secondary particles. Also shown are 90\% C.L. sensitivities of current (black; IceCube \citealt{2016PhRvL.117x1101A} and Auger \citealt{PhysRevD.91.092008}) and some future UHE neutrino detectors (grey; ARA/ARIANNA \citealt{2012APh....35..457A, 2015APh....70...12B}, GRAND \citealt{2017EPJWC.13502001M}, CHANT \citealt{2017PhRvD..95b3004N}). }
\end{figure}

The total flux contributed by all binary NS mergers over cosmological distances can be estimated by  
\begin{equation}
\Phi(E)= \frac{f_{\rm mag}}{4\pi} {\cal R}_0 \int\frac{c\,dz}{H(z)}\, f(z)\,\frac{d{N}}{dE'}(z),
\end{equation}
where ${\cal R}_0\sim 10^{-7}\,\rm Mpc^{-3}\,yr^{-1}$ is the estimated total rate\footnote{Although note that this rate is uncertain by at least 2 orders of magnitude, and current upper limits from Advanced LIGO allow a rate which is a factor of $\approx 10$ times higher than this fiducial estimate.} of NS binary mergers in the local universe (e.g.,~\citealt{2016ApJ...832L..21A}), $f_{\rm mag}$ is the fraction of mergers leaving long-lived or stable NS remnants, $H(z)$ is the Hubble constant at redshift z, and $f(z)$ describes the source evolution, which equals the ratio of the source rate at redshift z to that at today. Binary evolution models suggest that the NS-NS merger rate follows a history roughly comparable to the star formation rate (SFR) \citep{2013ApJ...779...72D}. Assuming that sources evolve with SFR \citep{0004-637X-613-1-200}, and taking the default parameters as in equation~\ref{eqn:fluence}, we obtain a peak flux of 
\bey
\left(E^2\Phi\right)_{\rm peak} &\approx& 2.8\times10^{-8}\,{\cal R}_{0,-7}f_{\rm mag}\,f_{\rm sup}^p \,f_{\rm sup}^\pi\,  f_{\rm sup}^\mu\,\\ \nonumber 
&&\rm GeV\,cm^{-2}\,s^{-1}\,sr^{-1}.  
\eey

Figure~\ref{fig:fluence_multi} presents the diffuse neutrino flux from numerical calculations, and demonstrates the flux dependence on the parameters $B$, $P_i$, and $M_{\rm ej}$.  We have assumed in all cases that $f_{\rm mag} = 1$ and vary one out of the three other parameters at a time. Compared to the default case, a stronger dipole magnetic field (faster magnetar spin-down rate) enhances the synchrotron cooling of cosmic rays and hence leads to lower neutrino production. If $B\gg 10^{15}$~G, the millisecond magnetar would be subjected to substantial gravitational wave losses, in addition to electromagnetic radiation considered in this work \citep{1969ApJ...157.1395O}. The spindown time would be too short to allow accelerated particles to leave the dense and highly magnetized nebula. As $t_{\rm p,rad}\propto B^{-2}$, particles would be quickly cooled by synchrotron radiation and not produce neutrinos. By contrast, a weaker dipole magnetic field allows a slower deposition of the spin-down energy into the nebula, such that most particles are injected in a later time when the radiative loss is less severe.  This also results in less interaction time for primary particles which results in a narrower distribution of the neutrino energy.  A magnetar with larger $P_i$ as might be produced if gravitational waves carry away a large fraction of the magnetar power, is less powerful to produce cosmic particles in general.  A higher mass ejecta expands less rapidly, making the cooling more efficient than decay for primary and secondary particles, and thus leads to a slightly lower neutrino flux at early times.

The diffuse flux of neutrinos from binary NS mergers for $f_{\rm mag} = 1$ and ${\cal R}_0\sim 10^{-7}\,\rm Mpc^{-3}\,yr^{-1}$ is consistent with the upper limit based on 7 years of IceCube data, $\Phi_{\rm UL}\sim 3\times 10^{-8}\,\rm GeV\,cm^{-2}\,s^{-1}\,sr^{-1}$ \citep{2016PhRvL.117x1101A}. The flux level is in a promising regime that can be detected by current and future experiments.  An absence of detection can in turn constrain the uncertain product $f_{\rm mag}\,{\cal R}_0$, or the magnetar fraction $f_{\rm mag}$ alone once the local merger rate ${\cal R}_0$ is measured by gravitational wave detectors.

\section{Discussion and Conclusions}\label{sec:discussion}

A long-lived, or indefinitely stable, millisecond magnetar may be formed by the coalescence of a binary NS system.  In the hours to days following the merger, the resulting powerful and high voltage pulsar wind inflates a magnetized nebula behind the merger ejecta in which particles can be accelerated up to ultra-high energies. Depending on the age of the source (time since merger), the accelerated cosmic rays are cooled by synchrotron emission in the nebula, or they may interact with the non-thermal and thermal radiation fields of nebula to produce high-energy neutrinos. 

Following the evolution of the radiative background and the interaction of cosmic rays and their secondary particles, we have explored the neutrino signatures of such magnetar remnants.  In optimistic scenarios in which an order unity fraction of NS mergers produce long-lived magnetar remnants with dipole magnetic field strengths of $\sim 10^{14}$~G, the cumulative neutrino background resulting from these events may be observed by the IceCube Observatory in the near future.  Even in less optimistic scenarios, in which the magnetar fraction is small or the dipole magnetic field is stronger, the diffuse flux is potentially within the reach of next-generation neutrino telescopes. 

GW information alone may not be sufficient to confirm or refute the presence of a long-lived magnetar, even in systems in which the chirp inspiral phase is detected with high SNR.  High frequency oscillations from the NS remnant (e.g.~\citealt{Clark+14}) can produce a measurable GW signal for hundreds of milliseconds following the merger and provide information on the NS EOS, but they are unlikely to provide unambiguous evidence for an extremely long remnant lifetime (the ringdown signature of the newly-formed BH will probably not be measurable).  The magnetar itself will produce a periodic gravitational wave signal; however, its strength depends on the presence of a strong toroidal magnetic field misaligned with the rotation axis (e.g., \citealt{Stella+05,2013PhRvD..88f7304F, 2014PhRvD..89d7302L, 2015ApJ...798...25D}) or the growth and saturation of the f-mode instability (e.g., \citealt{Doneva+15}).  

We propose that high energy neutrinos, with a characteristic light curve peaking days after the merger, could provide a comparatively ``clean" way to verify the presence of a long-lived magnetar.  Such a detection would also provide a more accurate sky localization for the source than provided by the GW signal, which could help identify the host galaxy if an electromagnetic counterpart is not detected.  Future neutrino telescopes, such as GRAND \citep{2017EPJWC.13502001M} and CHANT \citep{2017PhRvD..95b3004N}, are designed to improve the sensitivity of IceCube at EeVs by roughly two orders of magnitude. With such improved sensitivities, the magnetar neutrino  emission would be observable up to $\sim100$~Mpc, or at a rate of 0.4 NS merger event per year assuming ${\cal R}_0 = 10^{-7}\,\rm Mpc^{-3}\,yr^{-1}$ \citep{2016ApJ...832L..21A} and $f_{\rm mag}=1$.

% model dependence 
The fraction of NS mergers producing stable or long-lived magnetar remnants depends on the mass distribution of the merging NS binaries and, most sensitively, on the EoS of nuclear density matter through the maximum stable NS mass (e.g.~\citealt{Belczynski+08,Lawrence+15,Piro+17}). This is a large uncertainty of our model. But even in case that the maximum mass is relatively low $\lesssim 2.2\, M_\odot$, a significant amount of the rotational energy inherited from the merger can still be extracted from the merger remnant before it collapses into a black hole, in a timescale comparable to the spin-down time (see, e.g., \citealt{Metzger17}, their Fig.~8).  Cosmic rays accelerated before the collapse could still produce a neutrino signal in this case, though the neutrino light curve would decay more rapidly at late times than in the stable magnetar case.

% constraints from radio observations
Constraints have been placed on long-lived magnetar remnants of NS mergers from late-time radio observations of short-duration GRBs \citep{2014MNRAS.437.1821M, 2016ApJ...819L..22H, 2016ApJ...831..141F}.  These works performed searches for radio emission from a group of well-localized short-duration GRBs on a timescale of months to years after the bursts.  No coincident signal was found, deriving upper limits on the kinetic energy and the ejecta mass in several observational samples.   Similar constraints on radio transients from stable magnetars are, or will be, constrained also by past or future planned wide-field radio transient surveys (\citealt{Metzger+15c}).  While on the face these observations suggest $f_{\rm mag} \ll 1$, these constraints are sensitive to the assumed microphysical parameters of the shock and the density of the surrounding circumburst medium.   It is also important to keep in mind that mergers giving rise to magnetars instead of black holes may not produce detectable prompt gamma-ray  emission, in which case the GRB sample could be biased (though merger-produced magnetars should be accompanied by  luminous optical/X-ray counterparts; \citealt{2013ApJ...776L..40Y,MP14, Siegel&Ciolfi16a,Siegel&Ciolfi16b, 2017PhRvD..95f3016C}).  

% comparing to Piro & Kollmeier 
\citet{Piro&Kollmeier16} focused on the escape of UHECRs from low ejecta-mass explosions, suggesting stable magnetars from NS mergers as sources which can explain the rates and heavy composition of the UHECR measurement by the Auger Observatory \citep{Aab:2015bza}.  They also estimate the neutrino emission from the hadronic interaction between UHECRs and ejecta baryons.  However, \citet{Piro&Kollmeier16} ignored interaction with the radiation background, under the assumption that the cosmic rays are composed primarily of heavy nuclei,  in which case the neutrino signal is much weaker signal due to energy losses being dominated by photo-disintegration instead of pion creation.  

Our work instead focuses on the interaction of cosmic-rays with the evolving radiation background of the post-merger environment, motivated as follows.  First, cosmic rays in the energy range $\sim 10^{17.5}-10^{18.5}$~eV are inferred to possess a proton-dominated light composition by both the Auger Observatory \citep{Aab:2015bza} and the Telescope Array \citep{TA_ICRC15}.  The neutrino signal at energies of $10^{17}$~eV should therefore be strong if magnetars from NS mergers indeed contribute the bulk of the UHECRs.  Second, although the composition of the GRB jet and merger debris is likely to be composed of intermediate to heavy elements (resulting from nucleosynthesis following their decompression from high densities; \citealt{2011MNRAS.415.2495M}), the composition of the ions exacted from the stellar surface which are accelerated in the pulsar magnetosphere is less clear. Third, as shown in Figure~\ref{fig:t_proton}, the interaction rate of cosmic rays with photons exceeds by orders of magnitude that with the ejecta baryons. Thus, even if only a small fraction of the accelerated cosmic rays are protons, the photopion interaction will be more efficient than the hadronic interaction in producing neutrinos.  As a result, including the effects of the radiation background on the UHECR cooling results in different neutrino spectrum and light curve.  We also predict lower neutrino flux above $\sim 10^{18}$~eV and from highly magnetized systems (with $B\ge 10^{15}$~G) compared to \citet{Piro&Kollmeier16}.

% AIC, ultra-stripped envelope supernovae
Finally, we note that similar transients powered by NSs with ultra-strong magnetic fields may be formed from the accretion-induced collapses (AIC) of white dwarfs \citep{1976A&A....46..229C, 1991ApJ...367L..19N}, or possibly in iron core collapse supernovae with very low ejecta masses (so-called ``ultra-stripped envelope supernovae"; \citealp{2013ApJ...774...58D, 2014ApJ...794...23D, 2015MNRAS.451.2123T}). Such events might occur at a comparable or potentially higher rate than NS mergers, although their gravitational wave emission will be significantly weaker.  The neutrino emission from such events could be similar to that studied in this work, if these events also produce millisecond magnetars.

\acknowledgements
We thank Kohta Murase for helpful comments. 
KF acknowledges the support of a Joint Space-Science Institute prize postdoctoral fellowship.  
BDM gratefully acknowledges support from the National Science Foundation (AST-1410950, AST-1615084), NASA through the Astrophysics Theory Program (NNX16AB30G, NNX17AK43G) and the Fermi Guest Investigator Program (NNX16AR73G).  
\software{SOPHIA \citep{2000CoPhC.124..290M}}. 

\bibliography{MagnetarNeutrino} 

\begin{thebibliography}{}
\expandafter\ifx\csname natexlab\endcsname\relax\def\natexlab#1{#1}\fi
\providecommand{\url}[1]{\href{#1}{#1}}

\bibitem[{Aab {et~al.}(2015{\natexlab{a}})Aab, Abreu, Aglietta,
  {et~al.}}]{PhysRevD.91.092008}
Aab, A., Abreu, P., Aglietta, M., {et~al.} 2015{\natexlab{a}}, Phys. Rev. D,
  91, 092008.
\newblock \url{https://link.aps.org/doi/10.1103/PhysRevD.91.092008}

\bibitem[{Aab {et~al.}(2015{\natexlab{b}})}]{Aab:2015bza}
Aab, A., {et~al.} 2015{\natexlab{b}}, Proceedings of Science (ICRC2015),
  arXiv:1509.03732

\bibitem[{{Aartsen} {et~al.}(2014){Aartsen}, {Ackermann}, {Adams}, {Aguilar},
  {Ahlers}, {Ahrens}, {Altmann}, {Anderson}, {Arguelles}, {Arlen}, \&
  et~al.}]{2014PhRvD..90j2002A}
{Aartsen}, M.~G., {Ackermann}, M., {Adams}, J., {et~al.} 2014, \prd, 90, 102002

\bibitem[{Aartsen {et~al.}(2015)Aartsen, Ackermann, Adams, Aguilar, Ahlers,
  Ahrens, Altmann, Anderson, Archinger, Arguelles, Arlen, Auffenberg, Bai,
  Barwick, Baum, Bay, Baker, Beatty, Tjus, Becker, BenZvi, Berghaus, Berley,
  Bernardini, Bernhard, Besson, Binder, Bindig, Bissok, Blaufuss, Blumenthal,
  Boersma, Bohm, Bos, Bose, B{\"o}ser, Botner, Brayeur, Bretz, Brown, Buzinsky,
  Casey, Casier, Cheung, Chirkin, Christov, Christy, Clark, Classen,
  Clevermann, Coenders, Cowen, Silva, Daughhetee, Davis, Day, de~Andr{\'e},
  De~Clercq, Dembinski, De~Ridder, Desiati, de~Vries, de~Wasseige, de~With,
  DeYoung, D{\'\i}az-V{\'e}lez, Dumm, Dunkman, Eagan, Eberhardt, Ehrhardt,
  Eichmann, Eisch, Euler, Evenson, Fadiran, Fazely, Fedynitch, Feintzeig,
  Felde, Filimonov, Finley, Fischer-Wasels, Flis, Frantzen, Fuchs, Gaisser,
  Gaior, Gallagher, Gerhardt, Gier, Gladstone, Gl{\"u}senkamp, Goldschmidt,
  Golup, Gonzalez, Goodman, G{\'o}ra, Grant, Gretskov, Groh, Gro{\ss}, Ha,
  Haack, Ismail, Hallen, Hallgren, Halzen, Hanson, Hebecker, Heereman, Heinen,
  Helbing, Hellauer, Hellwig, Hickford, Hignight, Hill, Hoffman, Hoffmann,
  Homeier, Hoshina, Huang, Huelsnitz, Hulth, Hultqvist, In, Ishihara, Jacobi,
  Jacobsen, Japaridze, Jero, Jurkovic, Kaminsky, Kappes, Karg, Karle, Kauer,
  Keivani, Kelley, Kheirandish, Kiryluk, Kl{\"a}s, Klein, K{\"o}hne, Kohnen,
  Kolanoski, Koob, K{\"o}pke, Kopper, Kopper, Koskinen, Kowalski, Krings,
  Kroll, Kroll, Kunnen, Kurahashi, Kuwabara, Labare, Lanfranchi, Larsen,
  Larson, Lesiak-Bzdak, Leuermann, L{\"u}nemann, Madsen, Maggi, Mahn, Maruyama,
  Mase, Matis, Maunu, McNally, Meagher, Medici, Meli, Meures, Miarecki,
  Middell, Middlemas, Milke, Miller, Mohrmann, Montaruli, Morse, Nahnhauer,
  Naumann, Niederhausen, Nowicki, Nygren, Obertacke, Olivas, Omairat,
  O{\textquoteright}Murchadha, Palczewski, Paul, Pepper, de~los Heros,
  Pfendner, Pieloth, Pinat, Posselt, Price, Przybylski, P{\"u}tz, Quinnan,
  R{\"a}del, Rameez, Rawlins, Redl, Rees, Reimann, Relich, Resconi, Rhode,
  Richman, Riedel, Robertson, Rodrigues, Rongen, Rott, Ruhe, Ruzybayev,
  Ryckbosch, Saba, Sander, Sandroos, Santander, Sarkar, Schatto, Scheriau,
  Schmidt, Schmitz, Schoenen, Sch{\"o}neberg, Sch{\"o}nwald, Schukraft,
  Schulte, Schulz, Seckel, Sestayo, Seunarine, Shanidze, Smith, Soldin,
  Spiczak, Spiering, Stamatikos, Stanev, Stanisha, Stasik, Stezelberger,
  Stokstad, St{\"o}{\ss}l, Strahler, Str{\"o}m, Strotjohann, Sullivan,
  Sutherland, Taavola, Taboada, Tamburro, Ter-Antonyan, Terliuk, Te{\v
  s}i{\'c}, Tilav, Toale, Tobin, Tosi, Tselengidou, Unger, Usner, Vallecorsa,
  van Eijndhoven, Vandenbroucke, van Santen, Vanheule, Vehring, Voge, Vraeghe,
  Walck, Wallraff, Weaver, Wellons, Wendt, Westerhoff, Whelan, Whitehorn,
  Wichary, Wiebe, Wiebusch, Williams, Wissing, Wolf, Wood, Woschnagg, Xu, Xu,
  Xu, Yanez, Yodh, Yoshida, \& Zarzhit...}]{Aartsen:2015jx}
Aartsen, M.~G., Ackermann, M., Adams, J., {et~al.} 2015, The Astrophysical
  Journal, 807, 1

\bibitem[{{Aartsen} {et~al.}(2016){Aartsen}, {Abraham}, {Ackermann}, {Adams},
  {Aguilar}, {Ahlers}, {Ahrens}, {Altmann}, {Andeen}, {Anderson}, \&
  et~al.}]{2016PhRvL.117x1101A}
{Aartsen}, M.~G., {Abraham}, K., {Ackermann}, M., {et~al.} 2016, Physical
  Review Letters, 117, 241101

\bibitem[{{Aartsen} {et~al.}(2017){Aartsen}, {Abraham}, {Ackermann}, {Adams},
  {Aguilar}, {Ahlers}, {Ahrens}, {Altmann}, {Andeen}, {Anderson}, \&
  et~al.}]{2017ApJ...835..151A}
---. 2017, \apj, 835, 151

\bibitem[{{Abbott} {et~al.}(2016){Abbott}, {Abbott}, {Abbott}, {Abernathy},
  {Acernese}, {Ackley}, {Adams}, {Adams}, {Addesso}, {Adhikari}, \&
  et~al.}]{2016ApJ...832L..21A}
{Abbott}, B.~P., {Abbott}, R., {Abbott}, T.~D., {et~al.} 2016, \apjl, 832, L21

\bibitem[{{Abramovici} {et~al.}(1992){Abramovici}, {Althouse}, {Drever},
  {Gursel}, {Kawamura}, {Raab}, {Shoemaker}, {Sievers}, {Spero}, \&
  {Thorne}}]{1992Sci...256..325A}
{Abramovici}, A., {Althouse}, W.~E., {Drever}, R.~W.~P., {et~al.} 1992,
  Science, 256, 325

\bibitem[{Accadia {et~al.}(2012)Accadia, Acernese, Alshourbagy, Amico,
  Antonucci, Aoudia, Arnaud, Arnault, Arun, Astone, Avino, Babusci, Ballardin,
  Barone, Barrand, Barsotti, Barsuglia, Basti, Bauer, Beauville, Bebronne,
  Bejger, Beker, Bellachia, Belletoile, Beney, Bernardini, Bigotta, Bilhaut,
  Birindelli, Bitossi, Bizouard, Blom, Boccara, Boget, Bondu, Bonelli, Bonnand,
  Boschi, Bosi, Bouedo, Bouhou, Bozzi, Bracci, Braccini, Bradaschia, Branchesi,
  Briant, Brillet, Brisson, Brocco, Bulik, Bulten, Buskulic, Buy, Cagnoli,
  Calamai, Calloni, Campagna, Canuel, Carbognani, Carbone, Cavalier, Cavalieri,
  Cecchi, Cella, Cesarini, Chassande-Mottin, Chatterji, Chiche, Chincarini,
  Chiummo, Christensen, Clapson, Cleva, Coccia, Cohadon, Colacino, Colas,
  Colla, Colombini, Conforto, Corsi, Cortese, Cottone, Coulon, Cuoco,
  D'Antonio, Daguin, Dari, Dattilo, David, Davier, Day, Debreczeni, Carolis,
  Dehamme, Fabbro, Pozzo, del Prete, Derome, Rosa, DeSalvo, Dialinas, Fiore,
  Lieto, Emilio, Virgilio, Dietz, Doets, Dominici, Dominjon, Drago, Drezen,
  Dujardin, Dulach, Eder, Eleuteri, Enard, Evans, Fabbroni, Fafone, Fang,
  Ferrante, Fidecaro, Fiori, Flaminio, Forest, Forte, Fournier, Fournier,
  Franc, Francois, Frasca, Frasconi, Freise, Gaddi, Galimberti, Gammaitoni,
  Ganau, Garnier, Garufi, Gáspár, Gemme, Genin, Gennai, Gennaro, Giacobone,
  Giazotto, Giordano, Giordano, Girard, Gouaty, Grado, Granata, Granata, Grave,
  Greverie, Groenstege, Guidi, Hamdani, Hayau, Hebri, Heidmann, Heitmann,
  Hello, Hemming, Hennes, Hermel, Heusse, Holloway, Huet, Iannarelli,
  Jaranowski, Jehanno, Journet, Karkar, Ketel, Voet, Kovalik, Kowalska,
  Kreckelbergh, Krolak, Lacotte, Lagrange, Penna, Laval, Marec, Leroy,
  Letendre, Li, Lieunard, Liguori, Lodygensky, Lopez, Lorenzini, Loriette,
  Losurdo, Loupias, Mackowski, Maiani, Majorana, Magazzù, Maksimovic,
  Malvezzi, Man, Mancini, Mansoux, Mantovani, Marchesoni, Marion, Marin,
  Marque, Martelli, Masserot, Massonnet, Matone, Matone, Mazzoni, Menzinger,
  Michel, Milano, Minenkov, Mitra, Mohan, Montorio, Morand, Moreau, Moreau,
  Morgado, Morgia, Mosca, Moscatelli, Mours, Mugnier, Mul, Naticchioni, Neri,
  Nocera, Pacaud, Pagliaroli, Pai, Palladino, Palomba, Paoletti, Paoletti,
  Paoli, Pardi, Parguez, Parisi, Pasqualetti, Passaquieti, Passuello,
  Perciballi, Perniola, Persichetti, Petit, Pichot, Piergiovanni, Pietka,
  Pignard, Pinard, Poggiani, Popolizio, Pradier, Prato, Prodi, Punturo, Puppo,
  Qipiani, Rabaste, Rabeling, Rácz, Raffaelli, Rapagnani, Rapisarda, Re,
  Reboux, Regimbau, Reita, Remilleux, Ricci, Ricciardi, Richard, Ripepe,
  Robinet, Rocchi, Rolland, Romano, Rosińska, Roudier, Ruggi, Russo, Salconi,
  Sannibale, Sassolas, Sentenac, Solimeno, Sottile, Sperandio, Stanga, Sturani,
  Swinkels, Tacca, Taddei, Taffarello, Tarallo, Tissot, Toncelli, Tonelli,
  Torre, Tournefier, Travasso, Tremola, Turri, Vajente, van~den Brand, Broeck,
  van~der Putten, Vasuth, Vavoulidis, Vedovato, Verkindt, Vetrano, Véziant,
  Viceré, Vinet, Vilalte, Vitale, Vocca, Ward, Was, Yamamoto, Yvert, Zendri,
  \& Zhang}]{1748-0221-7-03-P03012}
Accadia, T., Acernese, F., Alshourbagy, M., {et~al.} 2012, Journal of
  Instrumentation, 7, P03012.
\newblock \url{http://stacks.iop.org/1748-0221/7/i=03/a=P03012}

\bibitem[{{Adri{\'a}n-Mart{\'{\i}}nez}
  {et~al.}(2016){Adri{\'a}n-Mart{\'{\i}}nez}, {Albert}, {Andr{\'e}},
  {Anghinolfi}, {Anton}, {Ardid}, {Aubert}, {Avgitas}, {Baret},
  {Barrios-Mart{\'{\i}}}, \& et~al.}]{2016PhRvD..93l2010A}
{Adri{\'a}n-Mart{\'{\i}}nez}, S., {Albert}, A., {Andr{\'e}}, M., {et~al.} 2016,
  \prd, 93, 122010

\bibitem[{{Albert} {et~al.}(2017){Albert}, {Andre}, {Anghinolfi}, {Anton},
  {Ardid}, {Aubert}, {Avgitas}, {Baret}, {Barrios-Marti}, {Basa}, \&
  et~al.}]{2017arXiv170306298A}
{Albert}, A., {Andre}, M., {Anghinolfi}, M., {et~al.} 2017, ArXiv e-prints,
  arXiv:1703.06298

\bibitem[{{Ara Collaboration} {et~al.}(2012){Ara Collaboration}, {Allison},
  {Auffenberg}, {Bard}, {Beatty}, {Besson}, {B{\"o}ser}, {Chen}, {Chen},
  {Connolly}, {Davies}, {Duvernois}, {Fox}, {Gorham}, {Grashorn}, {Hanson},
  {Haugen}, {Helbing}, {Hill}, {Hoffman}, {Hong}, {Huang}, {Huang}, {Ishihara},
  {Karle}, {Kennedy}, {Landsman}, {Liu}, {Macchiarulo}, {Mase}, {Meures},
  {Meyhandan}, {Miki}, {Morse}, {Newcomb}, {Nichol}, {Ratzlaff}, {Richman},
  {Ritter}, {Rott}, {Rotter}, {Sandstrom}, {Seckel}, {Touart}, {Varner},
  {Wang}, {Weaver}, {Wendorff}, {Yoshida}, \& {Young}}]{2012APh....35..457A}
{Ara Collaboration}, {Allison}, P., {Auffenberg}, J., {et~al.} 2012,
  Astroparticle Physics, 35, 457

\bibitem[{{Arons}(2003)}]{Arons03}
{Arons}, J. 2003, \apj, 589, 871

\bibitem[{{Baret} {et~al.}(2012){Baret}, {Bartos}, {Bouhou},
  {Chassande-Mottin}, {Corsi}, {Di Palma}, {Donzaud}, {Drago}, {Finley},
  {Jones}, {Klimenko}, {Kouchner}, {M{\'a}rka}, {M{\'a}rka}, {Moscoso}, {Papa},
  {Pradier}, {Prodi}, {Raffai}, {Re}, {Rollins}, {Salemi}, {Sutton}, {Tse},
  {Van Elewyck}, \& {Vedovato}}]{2012PhRvD..85j3004B}
{Baret}, B., {Bartos}, I., {Bouhou}, B., {et~al.} 2012, \prd, 85, 103004

\bibitem[{{Bartos} {et~al.}(2011){Bartos}, {Finley}, {Corsi}, \&
  {M{\'a}rka}}]{2011PhRvL.107y1101B}
{Bartos}, I., {Finley}, C., {Corsi}, A., \& {M{\'a}rka}, S. 2011, Physical
  Review Letters, 107, 251101

\bibitem[{{Barwick} {et~al.}(2015){Barwick}, {Berg}, {Besson}, {Binder},
  {Binns}, {Boersma}, {Bose}, {Braun}, {Buckley}, {Bugaev}, {Buitink},
  {Dookayka}, {Dowkontt}, {Duffin}, {Euler}, {Gerhardt}, {Gustafsson},
  {Hallgren}, {Hanson}, {Israel}, {Kiryluk}, {Klein}, {Kleinfelder},
  {Niederhausen}, {Olevitch}, {Persichelli}, {Ratzlaff}, {Rauch}, {Reed},
  {Roumi}, {Samanta}, {Simburger}, {Stezelberger}, {Tatar}, {Uggerhoj},
  {Walker}, {Yodh}, \& {Young}}]{2015APh....70...12B}
{Barwick}, S.~W., {Berg}, E.~C., {Besson}, D.~Z., {et~al.} 2015, Astroparticle
  Physics, 70, 12

\bibitem[{{Bauswein} \& {Stergioulas}(2017)}]{Bauswein&Stergioulas17}
{Bauswein}, A., \& {Stergioulas}, N. 2017, ArXiv e-prints, arXiv:1702.02567

\bibitem[{{Belczynski} {et~al.}(2008){Belczynski}, {O'Shaughnessy}, {Kalogera},
  {Rasio}, {Taam}, \& {Bulik}}]{Belczynski+08}
{Belczynski}, K., {O'Shaughnessy}, R., {Kalogera}, V., {et~al.} 2008,
  Astrophys. J. Lett., 680, L129

\bibitem[{{Berezinskii} {et~al.}(1990){Berezinskii}, {Bulanov}, {Dogiel}, \&
  {Ptuskin}}]{1990acr..book.....B}
{Berezinskii}, V.~S., {Bulanov}, S.~V., {Dogiel}, V.~A., \& {Ptuskin}, V.~S.
  1990, {Astrophysics of cosmic rays}

\bibitem[{{Blasi} {et~al.}(2000){Blasi}, {Epstein}, \&
  {Olinto}}]{2000ApJ...533L.123B}
{Blasi}, P., {Epstein}, R.~I., \& {Olinto}, A.~V. 2000, \apjl, 533, L123

\bibitem[{{Bucciantini} {et~al.}(2012){Bucciantini}, {Metzger}, {Thompson}, \&
  {Quataert}}]{2012MNRAS.419.1537B}
{Bucciantini}, N., {Metzger}, B.~D., {Thompson}, T.~A., \& {Quataert}, E. 2012,
  \mnras, 419, 1537

\bibitem[{{Canal} \& {Schatzman}(1976)}]{1976A&A....46..229C}
{Canal}, R., \& {Schatzman}, E. 1976, \aap, 46, 229

\bibitem[{{Cerutti} \& {Beloborodov}(2016)}]{2016SSRv..tmp...84C}
{Cerutti}, B., \& {Beloborodov}, A.~M. 2016, \ssr, arXiv:1611.04331

\bibitem[{{Cerutti} {et~al.}(2015){Cerutti}, {Philippov}, {Parfrey}, \&
  {Spitkovsky}}]{2015MNRAS.448..606C}
{Cerutti}, B., {Philippov}, A., {Parfrey}, K., \& {Spitkovsky}, A. 2015,
  \mnras, 448, 606

\bibitem[{Charles {et~al.}(2015)}]{TA_ICRC15}
Charles, J., {et~al.} 2015, Proceedings of Science (ICRC2015) 035.
\newblock
  \url{{https://pos.sissa.it/archive/conferences/236/035/ICRC2015_035.pdf}}

\bibitem[{{Chen} \& {Beloborodov}(2014)}]{2014ApJ...795L..22C}
{Chen}, A.~Y., \& {Beloborodov}, A.~M. 2014, \apjl, 795, L22

\bibitem[{{Cho} \& {Ryu}(2009)}]{2009ApJ...705L..90C}
{Cho}, J., \& {Ryu}, D. 2009, \apjl, 705, L90

\bibitem[{{Ciolfi} {et~al.}(2017{\natexlab{a}}){Ciolfi}, {Kastaun},
  {Giacomazzo}, {Endrizzi}, {Siegel}, \& {Perna}}]{Ciolfi+17}
{Ciolfi}, R., {Kastaun}, W., {Giacomazzo}, B., {et~al.} 2017{\natexlab{a}},
  \prd, 95, 063016

\bibitem[{{Ciolfi} {et~al.}(2017{\natexlab{b}}){Ciolfi}, {Kastaun},
  {Giacomazzo}, {Endrizzi}, {Siegel}, \& {Perna}}]{2017PhRvD..95f3016C}
---. 2017{\natexlab{b}}, \prd, 95, 063016

\bibitem[{{Clark} {et~al.}(2014){Clark}, {Bauswein}, {Cadonati}, {Janka},
  {Pankow}, \& {Stergioulas}}]{Clark+14}
{Clark}, J., {Bauswein}, A., {Cadonati}, L., {et~al.} 2014, \prd, 90, 062004

\bibitem[{{Dai} {et~al.}(2006){Dai}, {Wang}, {Wu}, \&
  {Zhang}}]{2006Sci...311.1127D}
{Dai}, Z.~G., {Wang}, X.~Y., {Wu}, X.~F., \& {Zhang}, B. 2006, Science, 311,
  1127

\bibitem[{{Dall'Osso} {et~al.}(2015){Dall'Osso}, {Giacomazzo}, {Perna}, \&
  {Stella}}]{2015ApJ...798...25D}
{Dall'Osso}, S., {Giacomazzo}, B., {Perna}, R., \& {Stella}, L. 2015, \apj,
  798, 25

\bibitem[{{Dominik} {et~al.}(2013){Dominik}, {Belczynski}, {Fryer}, {Holz},
  {Berti}, {Bulik}, {Mandel}, \& {O'Shaughnessy}}]{2013ApJ...779...72D}
{Dominik}, M., {Belczynski}, K., {Fryer}, C., {et~al.} 2013, \apj, 779, 72

\bibitem[{{Doneva} {et~al.}(2015){Doneva}, {Kokkotas}, \&
  {Pnigouras}}]{Doneva+15}
{Doneva}, D.~D., {Kokkotas}, K.~D., \& {Pnigouras}, P. 2015, Phys. Rev. D, 92,
  104040

\bibitem[{{Drout} {et~al.}(2013){Drout}, {Soderberg}, {Mazzali}, {Parrent},
  {Margutti}, {Milisavljevic}, {Sanders}, {Chornock}, {Foley}, {Kirshner},
  {Filippenko}, {Li}, {Brown}, {Cenko}, {Chakraborti}, {Challis}, {Friedman},
  {Ganeshalingam}, {Hicken}, {Jensen}, {Modjaz}, {Perets}, {Silverman}, \&
  {Wong}}]{2013ApJ...774...58D}
{Drout}, M.~R., {Soderberg}, A.~M., {Mazzali}, P.~A., {et~al.} 2013, \apj, 774,
  58

\bibitem[{{Drout} {et~al.}(2014){Drout}, {Chornock}, {Soderberg}, {Sanders},
  {McKinnon}, {Rest}, {Foley}, {Milisavljevic}, {Margutti}, {Berger},
  {Calkins}, {Fong}, {Gezari}, {Huber}, {Kankare}, {Kirshner}, {Leibler},
  {Lunnan}, {Mattila}, {Marion}, {Narayan}, {Riess}, {Roth}, {Scolnic},
  {Smartt}, {Tonry}, {Burgett}, {Chambers}, {Hodapp}, {Jedicke}, {Kaiser},
  {Magnier}, {Metcalfe}, {Morgan}, {Price}, \& {Waters}}]{2014ApJ...794...23D}
{Drout}, M.~R., {Chornock}, R., {Soderberg}, A.~M., {et~al.} 2014, \apj, 794,
  23

\bibitem[{{Duncan} \& {Thompson}(1992)}]{1992ApJ...392L...9D}
{Duncan}, R.~C., \& {Thompson}, C. 1992, \apjl, 392, L9

\bibitem[{{Eichler} {et~al.}(1989){Eichler}, {Livio}, {Piran}, \&
  {Schramm}}]{Eichler+89}
{Eichler}, D., {Livio}, M., {Piran}, T., \& {Schramm}, D.~N. 1989, Nature, 340,
  126

\bibitem[{{Eidelman} {et~al.}(2004){Eidelman}, {Hayes}, {Olive},
  {Aguilar-Benitez}, {Amsler}, {Asner}, {Babu}, {Barnett}, {Beringer},
  {Burchat}, {Carone}, {Caso}, {Conforto}, {Dahl}, {D'Ambrosio}, {Doser},
  {Feng}, {Gherghetta}, {Gibbons}, {Goodman}, {Grab}, {Groom}, {Gurtu},
  {Hagiwara}, {Hern{\'a}ndez-Rey}, {Hikasa}, {Honscheid}, {Jawahery}, {Kolda},
  {Kwon}, {Mangano}, {Manohar}, {March-Russell}, {Masoni}, {Miquel},
  {M{\"o}nig}, {Murayama}, {Nakamura}, {Navas}, {Pape}, {Patrignani}, {Piepke},
  {Raffelt}, {Roos}, {Tanabashi}, {Terning}, {T{\"o}rnqvist}, {Trippe},
  {Vogel}, {Wohl}, {Workman}, {Yao}, {Zyla}, {Armstrong}, {Gee}, {Harper},
  {Lugovsky}, {Lugovsky}, {Lugovsky}, {Rom}, {Artuso}, {Barberio}, {Battaglia},
  {Bichsel}, {Biebel}, {Bloch}, {Cahn}, {Casper}, {Cattai}, {Chivukula},
  {Cowan}, {Damour}, {Desler}, {Dobbs}, {Drees}, {Edwards}, {Edwards},
  {Elvira}, {Erler}, {Ezhela}, {Fetscher}, {Fields}, {Foster}, {Froidevaux},
  {Fukugita}, {Gaisser}, {Garren}, {Gerber}, {Gerbier}, {Gilman}, {Haber},
  {Hagmann}, {Hewett}, {Hinchliffe}, {Hogan}, {H{\"o}hler}, {Igo-Kemenes},
  {Jackson}, {Johnson}, {Karlen}, {Kayser}, {Kirkby}, {Klein}, {Kleinknecht},
  {Knowles}, {Kreitz}, {Kuyanov}, {Lahav}, {Langacker}, {Liddle}, {Littenberg},
  {Manley}, {Martin}, {Narain}, {Nason}, {Nir}, {Peacock}, {Quinn}, {Raby},
  {Ratcliff}, {Razuvaev}, {Renk}, {Rolandi}, {Ronan}, {Rosenberg}, {Sachrajda},
  {Sakai}, {Sanda}, {Sarkar}, {Schmitt}, {Schneider}, {Scott}, {Seligman},
  {Shaevitz}, {Sj{\"o}strand}, {Smoot}, {Spanier}, {Spieler}, {Spooner},
  {Srednicki}, {Stahl}, {Stanev}, {Suzuki}, {Tkachenko}, {Trilling},
  {Valencia}, {van Bibber}, {Vincter}, {Ward}, {Webber}, {Whalley},
  {Wolfenstein}, {Womersley}, {Woody}, {Zenin}, {Zhu}, \& {Particle Data
  Group}}]{2004PhLB..592....1E}
{Eidelman}, S., {Hayes}, K.~G., {Olive}, K.~A., {et~al.} 2004, Physics Letters
  B, 592, 1

\bibitem[{{Fan} {et~al.}(2013){Fan}, {Wu}, \& {Wei}}]{2013PhRvD..88f7304F}
{Fan}, Y.-Z., {Wu}, X.-F., \& {Wei}, D.-M. 2013, \prd, 88, 067304

\bibitem[{{Fang}(2015)}]{2015JCAP...06..004F}
{Fang}, K. 2015, \jcap, 6, 004

\bibitem[{{Fang} {et~al.}(2014){Fang}, {Kotera}, {Murase}, \&
  {Olinto}}]{2014PhRvD..90j3005F}
{Fang}, K., {Kotera}, K., {Murase}, K., \& {Olinto}, A.~V. 2014, \prd, 90,
  103005

\bibitem[{{Fang} {et~al.}(2012){Fang}, {Kotera}, \&
  {Olinto}}]{2012ApJ...750..118F}
{Fang}, K., {Kotera}, K., \& {Olinto}, A.~V. 2012, \apj, 750, 118

\bibitem[{{Fang} {et~al.}(2013){Fang}, {Kotera}, \&
  {Olinto}}]{2013JCAP...03..010F}
---. 2013, \jcap, 3, 010

\bibitem[{{Fern{\'a}ndez} \& {Metzger}(2013)}]{Fernandez&Metzger13}
{Fern{\'a}ndez}, R., \& {Metzger}, B.~D. 2013, Mon. Not. R. Astron. Soc., 435,
  502

\bibitem[{{Fern{\'a}ndez} \& {Metzger}(2016)}]{2016ARNPS..66...23F}
---. 2016, Annual Review of Nuclear and Particle Science, 66, 23

\bibitem[{{Fong} {et~al.}(2016){Fong}, {Metzger}, {Berger}, \&
  {{\"O}zel}}]{2016ApJ...831..141F}
{Fong}, W., {Metzger}, B.~D., {Berger}, E., \& {{\"O}zel}, F. 2016, \apj, 831,
  141

\bibitem[{{Freiburghaus} {et~al.}(1999){Freiburghaus}, {Rosswog}, \&
  {Thielemann}}]{Freiburghaus+99}
{Freiburghaus}, C., {Rosswog}, S., \& {Thielemann}, F. 1999, Astrophys. J.,
  525, L121

\bibitem[{{Fryer} {et~al.}(2015){Fryer}, {Belczynski}, {Ramirez-Ruiz},
  {Rosswog}, {Shen}, \& {Steiner}}]{2015ApJ...812...24F}
{Fryer}, C.~L., {Belczynski}, K., {Ramirez-Ruiz}, E., {et~al.} 2015, \apj, 812,
  24

\bibitem[{Gao {et~al.}(2013)Gao, Zhang, Wu, \& Dai}]{Gao:2013ka}
Gao, H., Zhang, B., Wu, X.-F., \& Dai, Z.-G. 2013, Physical Review D, 88,
  043010

\bibitem[{{Gao} \& {Fan}(2006)}]{Gao&Fan06}
{Gao}, W.-H., \& {Fan}, Y.-Z. 2006, Chinese Journal of Astronomy \&
  Astrophysics, 6, 513

\bibitem[{{Giacomazzo} \& {Perna}(2013)}]{2013ApJ...771L..26G}
{Giacomazzo}, B., \& {Perna}, R. 2013, \apjl, 771, L26

\bibitem[{{Giacomazzo} {et~al.}(2015){Giacomazzo}, {Zrake}, {Duffell},
  {MacFadyen}, \& {Perna}}]{2015ApJ...809...39G}
{Giacomazzo}, B., {Zrake}, J., {Duffell}, P.~C., {MacFadyen}, A.~I., \&
  {Perna}, R. 2015, \apj, 809, 39

\bibitem[{{Goldreich} \& {Julian}(1969)}]{Goldreich69}
{Goldreich}, P., \& {Julian}, W.~H. 1969, ApJ, 157, 869

\bibitem[{{Gompertz} {et~al.}(2015){Gompertz}, {van der Horst}, {O'Brien},
  {Wynn}, \& {Wiersema}}]{Gompertz+15}
{Gompertz}, B.~P., {van der Horst}, A.~J., {O'Brien}, P.~T., {Wynn}, G.~A., \&
  {Wiersema}, K. 2015, \mnras, 448, 629

\bibitem[{{Guilet} {et~al.}(2016){Guilet}, {Bauswein}, {Just}, \&
  {Janka}}]{Guilet+16}
{Guilet}, J., {Bauswein}, A., {Just}, O., \& {Janka}, H.-T. 2016, ArXiv
  e-prints, arXiv:1610.08532

\bibitem[{{Gunn} \& {Ostriker}(1969)}]{1969PhRvL..22..728G}
{Gunn}, J.~E., \& {Ostriker}, J.~P. 1969, Physical Review Letters, 22, 728

\bibitem[{Halzen(2016)}]{Halzen:2016ee}
Halzen, F. 2016, Nature Physics, 13, 232

\bibitem[{Harry \& the LIGO
  Scientific~Collaboration(2010)}]{0264-9381-27-8-084006}
Harry, G.~M., \& the LIGO Scientific~Collaboration. 2010, Classical and Quantum
  Gravity, 27, 084006.
\newblock \url{http://stacks.iop.org/0264-9381/27/i=8/a=084006}

\bibitem[{Hones \& Bergeson(1965)}]{JGR:JGR4198}
Hones, E.~W., \& Bergeson, J.~E. 1965, Journal of Geophysical Research, 70,
  4951.
\newblock \url{http://dx.doi.org/10.1029/JZ070i019p04951}

\bibitem[{{Horesh} {et~al.}(2016){Horesh}, {Hotokezaka}, {Piran}, {Nakar}, \&
  {Hancock}}]{2016ApJ...819L..22H}
{Horesh}, A., {Hotokezaka}, K., {Piran}, T., {Nakar}, E., \& {Hancock}, P.
  2016, \apjl, 819, L22

\bibitem[{{Hotokezaka} {et~al.}(2017){Hotokezaka}, {Kashiyama}, \&
  {Murase}}]{2017arXiv170406276H}
{Hotokezaka}, K., {Kashiyama}, K., \& {Murase}, K. 2017, ArXiv e-prints,
  arXiv:1704.06276

\bibitem[{{Hotokezaka} {et~al.}(2013){Hotokezaka}, {Kiuchi}, {Kyutoku},
  {Okawa}, {Sekiguchi}, {Shibata}, \& {Taniguchi}}]{Hotokezaka+13}
{Hotokezaka}, K., {Kiuchi}, K., {Kyutoku}, K., {et~al.} 2013, Phys. Rev. D, 87,
  024001

\bibitem[{{IceCube-Gen2 Collaboration} {et~al.}(2014){IceCube-Gen2
  Collaboration}, {:}, {Aartsen}, {Ackermann}, {Adams}, {Aguilar}, {Ahlers},
  {Ahrens}, {Altmann}, {Anderson}, \& et~al.}]{2014arXiv1412.5106I}
{IceCube-Gen2 Collaboration}, {:}, {Aartsen}, M.~G., {et~al.} 2014, ArXiv
  e-prints, arXiv:1412.5106

\bibitem[{{Just} {et~al.}(2015){Just}, {Bauswein}, {Pulpillo}, {Goriely}, \&
  {Janka}}]{Just+15}
{Just}, O., {Bauswein}, A., {Pulpillo}, R.~A., {Goriely}, S., \& {Janka}, H.-T.
  2015, Mon. Not. R. Astron. Soc., 448, 541

\bibitem[{{Kasen} {et~al.}(2015){Kasen}, {Fern{\'a}ndez}, \&
  {Metzger}}]{Kasen+15}
{Kasen}, D., {Fern{\'a}ndez}, R., \& {Metzger}, B.~D. 2015, Mon. Not. R.
  Astron. Soc., 450, 1777

\bibitem[{{Kennel} \& {Coroniti}(1984)}]{1984ApJ...283..694K}
{Kennel}, C.~F., \& {Coroniti}, F.~V. 1984, \apj, 283, 710

\bibitem[{{Kiuchi} {et~al.}(2015){Kiuchi}, {Sekiguchi}, {Kyutoku}, {Shibata},
  {Taniguchi}, \& {Wada}}]{Kiuchi+15}
{Kiuchi}, K., {Sekiguchi}, Y., {Kyutoku}, K., {et~al.} 2015, Phys. Rev. D, 92,
  064034

\bibitem[{{Kiziltan} {et~al.}(2013){Kiziltan}, {Kottas}, {De Yoreo}, \&
  {Thorsett}}]{2013ApJ...778...66K}
{Kiziltan}, B., {Kottas}, A., {De Yoreo}, M., \& {Thorsett}, S.~E. 2013, \apj,
  778, 66

\bibitem[{{Kotera} {et~al.}(2015){Kotera}, {Amato}, \&
  {Blasi}}]{2015JCAP...08..026K}
{Kotera}, K., {Amato}, E., \& {Blasi}, P. 2015, \jcap, 8, 026

\bibitem[{{Kotera} {et~al.}(2013){Kotera}, {Phinney}, \& {Olinto}}]{Kotera+13}
{Kotera}, K., {Phinney}, E.~S., \& {Olinto}, A.~V. 2013, \mnras, 432, 3228

\bibitem[{{Lasky} {et~al.}(2014){Lasky}, {Haskell}, {Ravi}, {Howell}, \&
  {Coward}}]{2014PhRvD..89d7302L}
{Lasky}, P.~D., {Haskell}, B., {Ravi}, V., {Howell}, E.~J., \& {Coward}, D.~M.
  2014, \prd, 89, 047302

\bibitem[{{Lattimer} \& {Schramm}(1976)}]{Lattimer&Schramm76}
{Lattimer}, J.~M., \& {Schramm}, D.~N. 1976, Astrophys. J., 210, 549

\bibitem[{{Lawrence} {et~al.}(2015){Lawrence}, {Tervala}, {Bedaque}, \&
  {Miller}}]{Lawrence+15}
{Lawrence}, S., {Tervala}, J.~G., {Bedaque}, P.~F., \& {Miller}, M.~C. 2015,
  \apj, 808, 186

\bibitem[{{Lemoine} {et~al.}(2015){Lemoine}, {Kotera}, \&
  {P{\'e}tri}}]{2015JCAP...07..016L}
{Lemoine}, M., {Kotera}, K., \& {P{\'e}tri}, J. 2015, \jcap, 7, 016

\bibitem[{Li {et~al.}(2008)Li, Lu, \& Li}]{0004-637X-682-2-1166}
Li, X.-H., Lu, F.-J., \& Li, Z. 2008, The Astrophysical Journal, 682, 1166.
\newblock \url{http://stacks.iop.org/0004-637X/682/i=2/a=1166}

\bibitem[{{Lippuner} {et~al.}(2017){Lippuner}, {Fern{\'a}ndez}, {Roberts},
  {Foucart}, {Kasen}, {Metzger}, \& {Ott}}]{Lippuner+17}
{Lippuner}, J., {Fern{\'a}ndez}, R., {Roberts}, L.~F., {et~al.} 2017, ArXiv
  e-prints, arXiv:1703.06216

\bibitem[{{Martineau-Huynh} {et~al.}(2017){Martineau-Huynh}, {Bustamante},
  {Carvalho}, {Charrier}, {De Jong}, {de Vries}, {Fang}, {Feng}, {Finley},
  {Gou}, {Gu}, {Hu}, {Kotera}, {Le Coz}, {Medina}, {Murase}, {Niess},
  {Oikonomou}, {Timmermans}, {Wang}, {Wu}, \& {Zhang}}]{2017EPJWC.13502001M}
{Martineau-Huynh}, O., {Bustamante}, M., {Carvalho}, W., {et~al.} 2017, in
  European Physical Journal Web of Conferences, Vol. 135, European Physical
  Journal Web of Conferences, 02001

\bibitem[{{Metzger}(2016)}]{2016arXiv161009381M}
{Metzger}, B.~D. 2016, ArXiv e-prints, arXiv:1610.09381

\bibitem[{{Metzger}(2017)}]{Metzger17}
---. 2017, Living Reviews in Relativity, 20, 3

\bibitem[{{Metzger} \& {Bower}(2014)}]{2014MNRAS.437.1821M}
{Metzger}, B.~D., \& {Bower}, G.~C. 2014, \mnras, 437, 1821

\bibitem[{{Metzger} \& {Fern{\'a}ndez}(2014)}]{Metzger&Fernandez14}
{Metzger}, B.~D., \& {Fern{\'a}ndez}, R. 2014, Mon. Not. R. Astron. Soc., 441,
  3444

\bibitem[{{Metzger} {et~al.}(2011){Metzger}, {Giannios}, \&
  {Horiuchi}}]{2011MNRAS.415.2495M}
{Metzger}, B.~D., {Giannios}, D., \& {Horiuchi}, S. 2011, \mnras, 415, 2495

\bibitem[{Metzger \& Piro(2014)}]{MP14}
Metzger, B.~D., \& Piro, A.~L. 2014, Monthly Notices of the Royal Astronomical
  Society, 439, 3916

\bibitem[{{Metzger} {et~al.}(2008){Metzger}, {Quataert}, \&
  {Thompson}}]{Metzger+08b}
{Metzger}, B.~D., {Quataert}, E., \& {Thompson}, T.~A. 2008, Mon. Not. R.
  Astron. Soc., 385, 1455

\bibitem[{{Metzger} {et~al.}(2014){Metzger}, {Vurm}, {Hasco{\"e}t}, \&
  {Beloborodov}}]{Metzger+14}
{Metzger}, B.~D., {Vurm}, I., {Hasco{\"e}t}, R., \& {Beloborodov}, A.~M. 2014,
  Mon. Not. R. Astron. Soc., 437, 703

\bibitem[{{Metzger} {et~al.}(2015){Metzger}, {Williams}, \&
  {Berger}}]{Metzger+15c}
{Metzger}, B.~D., {Williams}, P.~K.~G., \& {Berger}, E. 2015, Astrophys. J.,
  806, 224

\bibitem[{{M{\"o}sta} {et~al.}(2015){M{\"o}sta}, {Ott}, {Radice}, {Roberts},
  {Schnetter}, \& {Haas}}]{Mosta+15}
{M{\"o}sta}, P., {Ott}, C.~D., {Radice}, D., {et~al.} 2015, \nat, 528, 376

\bibitem[{{M{\"u}cke} {et~al.}(2000){M{\"u}cke}, {Engel}, {Rachen},
  {Protheroe}, \& {Stanev}}]{2000CoPhC.124..290M}
{M{\"u}cke}, A., {Engel}, R., {Rachen}, J.~P., {Protheroe}, R.~J., \& {Stanev},
  T. 2000, Computer Physics Communications, 124, 290

\bibitem[{{Murase}(2015)}]{2015AIPC.1666d0006M}
{Murase}, K. 2015, in American Institute of Physics Conference Series, Vol.
  1666

\bibitem[{{Murase} {et~al.}(2015){Murase}, {Kashiyama}, {Kiuchi}, \&
  {Bartos}}]{2015ApJ...805...82M}
{Murase}, K., {Kashiyama}, K., {Kiuchi}, K., \& {Bartos}, I. 2015, \apj, 805,
  82

\bibitem[{{Murase} {et~al.}(2009){Murase}, {M{\'e}sz{\'a}ros}, \&
  {Zhang}}]{Murase09}
{Murase}, K., {M{\'e}sz{\'a}ros}, P., \& {Zhang}, B. 2009, PRD, 79, 103001

\bibitem[{{Neronov} {et~al.}(2017){Neronov}, {Semikoz}, {Anchordoqui}, {Adams},
  \& {Olinto}}]{2017PhRvD..95b3004N}
{Neronov}, A., {Semikoz}, D.~V., {Anchordoqui}, L.~A., {Adams}, J.~H., \&
  {Olinto}, A.~V. 2017, \prd, 95, 023004

\bibitem[{{Nomoto} \& {Kondo}(1991)}]{1991ApJ...367L..19N}
{Nomoto}, K., \& {Kondo}, Y. 1991, \apjl, 367, L19

\bibitem[{{Ostriker} \& {Gunn}(1969)}]{1969ApJ...157.1395O}
{Ostriker}, J.~P., \& {Gunn}, J.~E. 1969, \apj, 157, 1395

\bibitem[{{{\"O}zel} {et~al.}(2010){{\"O}zel}, {Psaltis}, {Ransom}, {Demorest},
  \& {Alford}}]{2010ApJ...724L.199O}
{{\"O}zel}, F., {Psaltis}, D., {Ransom}, S., {Demorest}, P., \& {Alford}, M.
  2010, \apjl, 724, L199

\bibitem[{{Philippov} \& {Spitkovsky}(2014)}]{2014ApJ...785L..33P}
{Philippov}, A.~A., \& {Spitkovsky}, A. 2014, \apjl, 785, L33

\bibitem[{{Pinto} \& {Eastman}(2000)}]{2000ApJ...530..757P}
{Pinto}, P.~A., \& {Eastman}, R.~G. 2000, \apj, 530, 757

\bibitem[{{Piro} {et~al.}(2017){Piro}, {Giacomazzo}, \& {Perna}}]{Piro+17}
{Piro}, A.~L., {Giacomazzo}, B., \& {Perna}, R. 2017, ArXiv e-prints,
  arXiv:1704.08697

\bibitem[{{Piro} \& {Kollmeier}(2016{\natexlab{a}})}]{2016ApJ...826...97P}
{Piro}, A.~L., \& {Kollmeier}, J.~A. 2016{\natexlab{a}}, \apj, 826, 97

\bibitem[{{Piro} \& {Kollmeier}(2016{\natexlab{b}})}]{Piro&Kollmeier16}
---. 2016{\natexlab{b}}, \apj, 826, 97

\bibitem[{{Price} \& {Rosswog}(2006)}]{2006Sci...312..719P}
{Price}, D.~J., \& {Rosswog}, S. 2006, Science, 312, 719

\bibitem[{{Radice}(2017)}]{Radice17}
{Radice}, D. 2017, \apjl, 838, L2

\bibitem[{{Rowlinson} {et~al.}(2013){Rowlinson}, {O'Brien}, {Metzger},
  {Tanvir}, \& {Levan}}]{Rowlinson+13}
{Rowlinson}, A., {O'Brien}, P.~T., {Metzger}, B.~D., {Tanvir}, N.~R., \&
  {Levan}, A.~J. 2013, Mon. Not. R. Astron. Soc., 430, 1061

\bibitem[{{Rybicki} \& {Lightman}(1979)}]{1979rpa..book.....R}
{Rybicki}, G.~B., \& {Lightman}, A.~P. 1979, {Radiative processes in
  astrophysics}

\bibitem[{{Sekiguchi} {et~al.}(2016){Sekiguchi}, {Kiuchi}, {Kyutoku},
  {Shibata}, \& {Taniguchi}}]{Sekiguchi+16}
{Sekiguchi}, Y., {Kiuchi}, K., {Kyutoku}, K., {Shibata}, M., \& {Taniguchi}, K.
  2016, \prd, 93, 124046

\bibitem[{{Shibata} \& {Taniguchi}(2006)}]{Shibata&Taniguchi06}
{Shibata}, M., \& {Taniguchi}, K. 2006, Phys. Rev. D, 73, 064027

\bibitem[{{Siegel} \& {Ciolfi}(2016{\natexlab{a}})}]{Siegel&Ciolfi16a}
{Siegel}, D.~M., \& {Ciolfi}, R. 2016{\natexlab{a}}, Astrophys. J., 819, 14

\bibitem[{{Siegel} \& {Ciolfi}(2016{\natexlab{b}})}]{Siegel&Ciolfi16b}
---. 2016{\natexlab{b}}, Astrophys. J., 819, 15

\bibitem[{{Siegel} {et~al.}(2013){Siegel}, {Ciolfi}, {Harte}, \&
  {Rezzolla}}]{Siegel+13}
{Siegel}, D.~M., {Ciolfi}, R., {Harte}, A.~I., \& {Rezzolla}, L. 2013, Phys.
  Rev. D, 87, 121302

\bibitem[{{Siegel} \& {Metzger}(2017)}]{Siegel&Metzger17}
{Siegel}, D.~M., \& {Metzger}, B.~D. 2017, ArXiv e-prints, arXiv:1705.05473

\bibitem[{{Spitkovsky}(2006)}]{Spitkovsky06}
{Spitkovsky}, A. 2006, Astrophys. J. Lett., 648, L51

\bibitem[{{Stella} {et~al.}(2005){Stella}, {Dall'Osso}, {Israel}, \&
  {Vecchio}}]{Stella+05}
{Stella}, L., {Dall'Osso}, S., {Israel}, G.~L., \& {Vecchio}, A. 2005, \apjl,
  634, L165

\bibitem[{Strolger {et~al.}(2004)Strolger, Riess, Dahlen, Livio, Panagia,
  Challis, Tonry, Filippenko, Chornock, Ferguson, Koekemoer, Mobasher,
  Dickinson, Giavalisco, Casertano, Hook, Bondin, Leibundgut, Nonino, Rosati,
  Spinrad, Steidel, Stern, Garnavich, Matheson, Grogin, Hornschemeier,
  Kretchmer, Laidler, Lee, Lucas, de~Mello, Moustakas, Ravindranath,
  Richardson, \& Taylor}]{0004-637X-613-1-200}
Strolger, L.-G., Riess, A.~G., Dahlen, T., {et~al.} 2004, The Astrophysical
  Journal, 613, 200.
\newblock \url{http://stacks.iop.org/0004-637X/613/i=1/a=200}

\bibitem[{{Svensson}(1987)}]{1987MNRAS.227..403S}
{Svensson}, R. 1987, \mnras, 227, 403

\bibitem[{{Symbalisty} \& {Schramm}(1982)}]{Symbalisty&Schramm82}
{Symbalisty}, E., \& {Schramm}, D.~N. 1982, Astrophys. J. Lett., 22, 143

\bibitem[{{Tauris} {et~al.}(2015){Tauris}, {Langer}, \&
  {Podsiadlowski}}]{2015MNRAS.451.2123T}
{Tauris}, T.~M., {Langer}, N., \& {Podsiadlowski}, P. 2015, \mnras, 451, 2123

\bibitem[{{The IceCube Collaboration}(2013)}]{1242856}
{The IceCube Collaboration}. 2013, Science, 342,
  http://science.sciencemag.org/content/342/6161/1242856.full.pdf.
\newblock \url{http://science.sciencemag.org/content/342/6161/1242856}

\bibitem[{{The IceCube Collaboration} {et~al.}(2015){The IceCube
  Collaboration}, {Aartsen}, {Abraham}, {Ackermann}, {Adams}, {Aguilar},
  {Ahlers}, {Ahrens}, {Altmann}, {Anderson}, \& et~al.}]{2015arXiv151005223T}
{The IceCube Collaboration}, {Aartsen}, M.~G., {Abraham}, K., {et~al.} 2015,
  ICRC 2015 proceedings, arXiv:1510.05223

\bibitem[{{Thompson} {et~al.}(2004){Thompson}, {Chang}, \&
  {Quataert}}]{Thompson+04}
{Thompson}, T.~A., {Chang}, P., \& {Quataert}, E. 2004, Astrophys. J., 611, 380

\bibitem[{{Wollaeger} {et~al.}(2017){Wollaeger}, {Korobkin}, {Fontes},
  {Rosswog}, {Even}, {Fryer}, {Sollerman}, {Hungerford}, {van Rossum}, \&
  {Wollaber}}]{Wollaeger+17}
{Wollaeger}, R.~T., {Korobkin}, O., {Fontes}, C.~J., {et~al.} 2017, ArXiv
  e-prints, arXiv:1705.07084

\bibitem[{{Wu} {et~al.}(2016){Wu}, {Fern{\'a}ndez}, {Mart{\'{\i}}nez-Pinedo},
  \& {Metzger}}]{Wu+16}
{Wu}, M.-R., {Fern{\'a}ndez}, R., {Mart{\'{\i}}nez-Pinedo}, G., \& {Metzger},
  B.~D. 2016, Mon. Not. R. Astron. Soc., arXiv:1607.05290

\bibitem[{{Yu} {et~al.}(2013){Yu}, {Zhang}, \& {Gao}}]{2013ApJ...776L..40Y}
{Yu}, Y.-W., {Zhang}, B., \& {Gao}, H. 2013, \apjl, 776, L40

\bibitem[{{Zrake} \& {MacFadyen}(2013)}]{2013ApJ...769L..29Z}
{Zrake}, J., \& {MacFadyen}, A.~I. 2013, \apjl, 769, L29

\end{thebibliography}

  \end{document}